\documentclass[article,onecolumn]{revtex4}
\usepackage{makeidx}
\usepackage{amssymb}
\usepackage{amsmath}
\usepackage{graphicx}
\usepackage{dcolumn}
\usepackage{bm}
\usepackage[center]{subfigure}
\usepackage{color}
\usepackage[bf, small]{titlesec}
\usepackage[symbol]{footmisc}

\begin{document}

\title{Singular Mean-Field States: a Brief Review of Recent Results}
\author{Elad Shamriz, Zhaopin Chen, and Boris A. Malomed$^{1,2,3}$}
\affiliation{$^{1}$Department of Physical Electronics, School of Electrical Engineering,
Faculty of Engineering, Tel Aviv University, Tel Aviv 69978, Israel\\
$^{2}$Center for Light-Matter Interaction, Tel Aviv University, Tel Aviv
University, Tel Aviv 69978, Israel\\
$^{3}$malomed@tauex.tau.ac.il}
\author{Hidetsugu Sakaguchi}
\affiliation{Department of Applied Science for Electronics and Materials,
Interdisciplinary Graduate School of Engineering Sciences,}
\affiliation{Kyushu University, Kasuga, Fukuoka 816-8580, Japan}

\begin{abstract}
This article provides a focused review of recent findings which demonstrate,
in some cases quite counter-intuitively, the existence of bound states with
a singularity of the density pattern at the center, while the states are
physically meaningful because their total norm converges. One model of this
type is based on the 2D Gross-Pitaevskii equation (GPE) which combines the
attractive potential $\sim r^{-2}$ and the quartic self-repulsive
nonlinearity, induced by the Lee-Huang-Yang effect (quantum fluctuations
around the mean-field state). The GPE demonstrates suppression of the 2D
quantum collapse, driven by the attractive potential, and emergence of a
stable ground state (GS), whose density features an integrable singularity $\sim r^{-4/3}$ at $r\rightarrow 0$. 
Modes with embedded angular momentum exist too, and they have their stability region. A counter-intuitive peculiarity of the
model is that the GS exists even if the sign of the potential is reversed
from attraction to repulsion, provided that its strength is small enough.
This peculiarity finds a relevant explanation. The other model outlined in
the review includes 1D, 2D, and 3D GPEs, with the septimal (seventh-order),
quintic, and cubic self-repulsive terms, respectively. These equations give
rise to stable singular solitons, which represent the GS for each dimension $D$, with the density singularity $\sim r^{-2/(4-D)}$. Such states may be considered as a result of screening of a ``bare"
delta-functional attractive potential by the respective nonlinearity.

\noindent
This paper was published in:

\noindent
Condensed Matter \textbf{5}, 20 (2020); https://doi.org/10.3390/condmat5010020

\end{abstract}
\maketitle

\noindent
The list of acronyms:

\noindent
\begin{tabular}{@{}ll}
2D & two-dimensional \\
3D & three-dimensional \\
BEC & Bose-Einstein condensate \\
GPE & Gross-Pitaevskii equation \\
GS & ground state \\
LHY & Lee-Huang-Yang (correction to the mean-field theory) \\
MF & mean field \\
NLSE & nonlinear Schr\"{o}dinger equation \\
TF & Thomas-Fermi (approximation) \\
VK & Vakhitov-Kolokolov (stability criterion)
\end{tabular}

\section{Introduction}

\subsection{Singular states pulled to the center by attractive fields}

\subsubsection{Outline of the topic}

A commonly adopted condition which distinguishes physically relevant states
produced by various two- and three-dimensional (2D and 3D) models
originating in quantum mechanics, studies of Bose-Einstein condensates
(BECs), nonlinear optics, plasma physics, etc., is that the respective
fields, such as wave functions in quantum mechanics and mean-field (MF)
description of BEC, or local amplitudes of optical fields, must avoid
singularities at $r\rightarrow 0$, where $r$ is the radial coordinate. A
usual example of the relevance of this condition is provided by the
quantum-mechanical Schr\"{o}dinger equation with the attractive Coulomb
potential, $\mathcal{U}(r)\sim -r^{-1}$: the stationary real wave function
of states with integer azimuthal quantum number (alias vorticity) $l\geq 0$
has expansion
\begin{equation}
u(r)=\left[ u_{0}+u_{1}r+\mathcal{O}\left( r^{2}\right) \right] r^{l}
\label{l}
\end{equation}%
at $r\rightarrow 0,$ thus avoiding a singularity, despite the fact that the
trapping potential is singular \cite{LL}.

In quantum mechanics, a critical role is played by a more singular
attractive potential, \textit{viz}.,
\begin{equation}
\mathcal{U}(r)=-\left( U_{0}/2\right) r^{-2},  \label{U(r)}
\end{equation}%
with $U_{0}>0$, which gives rise to the \textit{quantum collapse}, alias
``\textit{fall onto the center}" \cite{LL}. This well-known
phenomenon means nonexistence of the ground state (GS) in the 3D and 2D Schr%
\"{o}dinger equations with potential (\ref{U(r)}). In 3D, the collapse
occurs when $U_{0}$ exceeds a finite critical value, $\left( U_{0}^{\mathrm{%
(3D)}}\right) _{\mathrm{cr}}$ (in the notation adopted below in Eq. (\ref%
{3Dscaled}), $\left( U_{0}^{\mathrm{(3D)}}\right) _{\mathrm{cr}}=1/4$),
while in two dimensions $\left( U_{0}^{\mathrm{(2D)}}\right) _{\mathrm{cr}%
}=0 $, i.e., the 2D collapse happens at any $U_{0}>0$.

In both 3D and 2D cases, potential (\ref{U(r)}) may be realized as
electrostatic pull of a particle (small molecule), carrying a permanent
electric dipole moment, to a charge placed at the center, assuming that the
local orientation of the dipole is fixed by the minimization of its energy
in the central field \cite{HS1}. In addition to that, in the 2D case the
same potential (\ref{U}) may be realized as attraction of a magnetically
polarizable atom to a thread carrying electric current (e.g., an electron
beam) transversely to the system's plane, or the attraction of an
electrically polarizable atom to a uniformly charged transverse thread
(other 2D settings in Bose-Einstein condensates (BECs) under the action of
similar fields were considered in Refs. \cite{Austria1} and \cite{Austria2}).

A fundamental issue is stabilization of the 3D and 2D quantum-mechanical
settings with pulling potential (\ref{U(r)}) against the collapse, with the
aim to create a missing GS. A solution was proposed in Refs. \cite{QTF1}-%
\cite{QTF3}, which replaced the original quantum-mechanical problem by one
based on a linear quantum-field theory. While such a model produces a GS, it
does not answer a natural question: what an effective radius of the GS is
for given parameters of the setting, such as $U_{0}$ in (\ref{U(r)}) and the
mass of the quantum particle, $m$. Actually, the field-theory solution
defines the GS size as an arbitrary spatial scale, which varies as a
parameter of the respective field-theory renormalization group. A related
problem is the definition of self-adjoint Hamiltonians in the linear quantum
theory including the interaction of a particle with singular potentials \cite%
{Yafaev,Noja}.

Another solution was proposed in Ref. \cite{HS1}, which replaced the 3D
linear Schr\"{o}dinger equation by a nonlinear Gross-Pitaevskii equation
(GPE) \cite{GP} for a gas of particles attracted to the center by potential (%
\ref{U}), with repulsive inter-particle collisions. In the framework of the
mean-field (MF)\ approximation, the 3D GPE gives rise to the missing GS at
all values of $U_{0}>\left( U_{0}^{\mathrm{(3D)}}\right) _{\mathrm{cr}}$.
The radius of the GS is fully determined by model's constants, i.e., $U_{0}$%
, $m$, the scattering length of the inter-particle collisions, and the
number of particles, $N$. For typical values of the physical parameters, an
estimate for the radius is a few microns. Beyond the framework of the MF, it
was demonstrated that the many-body quantum theory, applied to the same
setting, does not, strictly speaking, create GS, but the interplay of the
attraction to the center and inter-particle repulsion gives rise to a
metastable state. For sufficiently large $N$, the metastable state is nearly
tantamount to GS, being separated from the collapse regime by a very tall
potential barrier \cite{Gregory}. Further, the mean-field GS was also
constructed in the 3D gas embedded in a strong uniform electric field, which
reduces the symmetry of the effective pulling potential from spherical to
cylindrical \cite{HS2}, as well as in the two-component 3D\ gas \cite{HS3}.

In 2D, the problem is more difficult, as the repulsive cubic term in GPE,
which represents inter-atomic collisions in the MF approximation \cite{GP},
is not strong enough to suppress the 2D quantum collapse and create the GS.
The main issue is that, in 3D and 2D settings alike, the MF wave function, $%
\psi (r)$, produced by the respective GPE, has density $\left\vert \psi
(r)\right\vert ^{2}$ diverging $\sim r^{-2}$ at $r\rightarrow 0$. In terms
of the integral norm,%
\begin{equation}
N=\left( 2\pi \right) ^{D-1}\int_{0}^{\infty }\left\vert \psi (r)\right\vert
^{2}r^{D-1}dr  \label{N}
\end{equation}%
($D=3$ or $2$ is the dimension), the density singularity $r^{-2}$ is
integrable in 3D, but not in 2D, where it gives rise to a logarithmic
divergence of the norm:%
\begin{equation}
N\sim \ln \left( r_{\mathrm{cutoff}}^{-1}\right) ,  \label{ln()}
\end{equation}%
where $r_{\mathrm{cutoff}}$ is a cutoff (smallest) radius, which may be
determined by the size of particles in the condensate. Actually, the cubic
self-repulsion is \emph{critical} in 2D, as any stronger nonlinear term is
sufficient to stabilize the 2D setting. In 3D, the critical value of the
repulsive-nonlinearity power, which also leads to the logarithmic divergence
of $N$, is $7/3$; it is relevant to mention that the respective nonlinear
term, $|\psi |^{4/3}\psi $, represents the effective repulsion in the
density-functional model of the Fermi gas \cite{Fermi2}-\cite{Fermi3}.

A solution for the 2D setting is offered by the quintic defocusing
nonlinearity \cite{HS1}, that may account for three-body repulsive
collisions in the bosonic gas \cite{Abdullaev1,Abdullaev2}. However, a
difficulty in the physical realization of the quintic term is the fact that
three-body collisions give rise to effective losses, kicking out particles
from BEC to the thermal component of the gas \cite{loss1}-\cite{loss3}.

In addition to original papers \cite{HS1,HS2,HS3} and \cite{Gregory}, the
above-mentioned results were summarized in short review \cite{CM}.

\subsubsection{New results included in the review}

Recently, much interest was drawn to quasi-2D and 3D self-trapped states in
BEC in the form of ``quantum droplets", filled by a nearly
incompressible two-component condensate, which is considered as an
ultradilute quantum fluid. This possibility was predicted in the framework
of the 3D \cite{Petrov} and 2D \cite{Petrov-Astra,Zin,Santos} GPEs which
include the Lee-Huang-Yang (LHY) corrections to the MF approximation \cite%
{LHY}. They represent effects of quantum fluctuations around the MF states.
The two-component structure of the condensate makes it possible to provide
nearly complete cancellation between the inter-component MF attraction and
intra-component repulsion (which, in turn, can be adjusted by means of the
Feshbach-resonance technique \cite{Feshbach})\ and thus create stable
droplets through the balance of the relatively weak residual MF attraction
and LHY-induced quartic self-repulsion. The quantum droplets of oblate
(quasi-2D) \cite{Leticia1,Leticia2} and fully 3D (isotropic) \cite%
{Inguscio1,Inguscio2} shapes were created in a mixture of two different spin
states of $^{39}$K atoms, as well as in a mixture of $^{41}$K and $^{87}$Rb
atoms \cite{hetero}. Further, it was predicted that 2D \cite%
{Raymond1,Raymond2,lattice} and 3D \cite{Barcelona} droplets with \emph{%
embedded vorticity} have their stability regions too. The LHY effect has
also opened the way to the creation of stable 3D droplets in
single-component BEC with long-range interactions between atoms carrying
magnetic dipole moments \cite{Pfau1}-\cite{Pfau5}, although
dipolar-condensate droplets with embedded vorticity are unstable \cite{Macri}%
.

The LHY effect in the 3D\ model of BEC pulled to the center by potential (%
\ref{U(r)}) was considered in Ref. \cite{CM}, with a conclusion that the LHY
term gives rise to a quantum phase transition at $U_{0}=2/9$ (note that it
is close to but smaller than the above-mentioned critical value for the
linear Schr\"{o}dinger equation, $\left( U_{0}^{\mathrm{(3D)}}\right) _{%
\mathrm{cr}}=1/4$). The phase transition manifests itself in the change of
the asymptotic form of the GS stationary wave function at $r\rightarrow 0$:%
\begin{equation}
u(r)\approx \left\{
\begin{array}{c}
\mathrm{const}\cdot r^{-\left( 1/2-\sqrt{1/4-U_{0}}\right) },~\mathrm{at}%
~U_{0}<2/9, \\
\left( U_{0}/2-1/9\right) ^{1/3}r^{-2/3},~\mathrm{at}~U_{0}>2/9.%
\end{array}%
\right.  \label{transition}
\end{equation}%
This phase transition may be categorized as one of the first kind, as power $%
\alpha $ in terms $r^{-\alpha }$ in Eq. (\ref{transition}) undergoes a
finite jump at $U_{0}=2/9$, from $1/3$ to $2/3$.

In Section 2 of the present review we summarize recent results which
demonstrate that the stabilization of the GS in the quasi-2D bosonic gas
pulled to the center by potential (\ref{U}) may be provided by the LHY
correction to the GPE \cite{we}. This possibility is essential because, as
mentioned above, the alternative, in the form of the quintic self-repulsion,
is problematic in the BEC setting. The underlying three-dimensional GPE,
which includes the LHY quartic defocusing term, is%
\begin{equation}
i\hbar \frac{\partial \Psi }{\partial t}=-\frac{\hbar ^{2}}{2m}\nabla
^{2}\Psi +W(\mathbf{r})\Psi +\frac{4\pi \hbar ^{2}\delta a}{m}|\Psi
|^{2}\Psi +\frac{256\sqrt{2\pi }\hbar ^{2}}{3m}a^{5/2}|\Psi |^{3}\Psi ,
\label{3DGPE}
\end{equation}%
where $\Psi $ stands for equal wave functions of two components of the BEC, $%
W(\mathbf{r})$ is the general trapping potential, $a>0$ is the scattering
length of inter-particle collisions, $\delta a\gtrless 0$, with $\left\vert
\delta a\right\vert \ll a$, represents the above-mentioned small disbalance
of the inter-component attraction and intra-component repulsion, and the
last term in Eq. (\ref{3DGPE}) is the LHY correction to the MF equation \cite%
{Petrov}.

The reduction of Eq. (\ref{3DGPE}) to the 2D form, with coordinates $\left(
x,y\right) $, under the action of tight confinement applied in the $z$
direction, was derived in Ref. \cite{Petrov-Astra}, producing the GPE with a
cubic term multiplied by an additional logarithmic factor,
\begin{equation}
\left( \mathrm{nonlin}\right) _{\mathrm{2D}}\sim |\Psi |^{2}\ln \left( |\Psi
|^{2}/\Psi _{0}^{2}\right) \Psi .  \label{ln}
\end{equation}%
However, this limit implies extremely strong confinement in the $z$
direction, with the transverse size $a_{\perp }\ll \xi $, where the healing
length is $\xi =\left( 32\sqrt{2}/3\pi \right) \left( a/\left\vert \delta
a\right\vert \right) ^{3/2}a\approx 5\left( a/\left\vert \delta a\right\vert
\right) ^{3/2}a$ \cite{Petrov}. For experimentally relevant parameters \cite%
{Leticia1}-\cite{Inguscio2}, an estimate is $\xi \simeq 30$ nm. On the other
hand, a realistic size of the confinement length in the experiment is few $%
\mathrm{\mu }$m, implying relation $a_{\perp }\gg \xi $, opposite to the
above-mentioned one necessary for the derivation of Eq. (\ref{ln}).
Therefore, it is relevant to reduce Eq. (\ref{3DGPE}) to the 2D form,
keeping the same nonlinearity as in Eq. (\ref{3DGPE}).

To complete the derivation of the effective 2D equation, we first rescale
three-dimensional Eq. (\ref{3DGPE}), measuring the density, length, time,
and the trapping potential in units of $\left( 36/25\right) n_{0}$, $\xi $, $%
\tau \equiv \left( m/\hbar \right) \xi ^{2}$, and $\hbar /\tau $,
respectively:%
\begin{equation}
i\frac{\partial \Psi }{\partial t}=-\frac{1}{2}\nabla ^{2}\Psi +\sigma |\Psi
|^{2}\Psi +|\Psi |^{3}\Psi +W(\mathbf{r})\Psi ,  \label{3Dscaled}
\end{equation}%
where $\sigma =\pm 1$ is the sign of $\delta a$, the potential is a sum of
term (\ref{U(r)}) and a transverse-confinement term, $\left( 1/2\right)
a_{\perp }^{-4}z^{2}$, with sufficiently small $a_{\perp }^{2}$. Then, the
3D $\rightarrow $ 2D reduction is performed by means of the usual
substitution \cite{Luca,Delgado}, $\Psi \left( x,y,z,t\right) =\psi \left(
x,y,t\right) \exp \left( -z^{2}/2a_{\perp }^{2}\right) $, followed by
averaging in the transverse direction, $z$. Additional rescaling, $\psi
\rightarrow \left( 2/\sqrt{5}\right) \psi $, $\left( x,y\right) \rightarrow
\left( \sqrt{5}/2\right) \left( x,y\right) $, and $t\rightarrow \left(
5/4\right) t$, casts the effective 2D equation, written in polar coordinates
$\left( r,\theta \right) $, in the final form:
\begin{equation}
i\frac{\partial \psi }{\partial t}=-\frac{1}{2}\left( \frac{\partial
^{2}\psi }{\partial r^{2}}+\frac{1}{r}\frac{\partial \psi }{\partial r}+%
\frac{1}{r^{2}}\frac{\partial ^{2}\psi }{\partial \theta ^{2}}\right) -\frac{%
U_{0}}{2r^{2}}\psi +\sigma \left\vert \psi \right\vert ^{2}\psi +|\psi
|^{3}\psi ,  \label{psi2d}
\end{equation}%
which includes potential (\ref{U(r)}).

In the framework of Eq. (\ref{psi2d}), it is also relevant to consider the
case of $\delta a=0$, which implies exact compensation of the
inter-component attraction and intra-component repulsion. In this case, one
should set $\sigma =0$ in Eq. (\ref{psi2d}), keeping the nonlinearity which
originates as the LHY correction to the MF field theory (cf. Ref. \cite%
{LHY-only}).

Results for both GS and vortex states, produced by the analysis of the 2D
equation (\ref{psi2d}) \cite{we}, are summarized, as a part of the present
review, in Section 2. Essential conclusions are that all the GS solutions
(with zero vorticity) are stable, while vortex modes are stable if 
$U_0$ exceeds a certain critical value, which depends on the vorticity.

\subsection{Singular solitons: previously known results}

\subsubsection{Singular solitons in free space}

The condition of the convergence of the integral norm is equally relevant
for self-trapped states, which are predicted, as localized solutions, by
models such as the nonlinear Schr\"{o}dinger equation (NLSE) for a complex
wave function, $\psi \left( x,y,z,t\right) $:%
\begin{equation}
i\psi _{t}+(1/2)\nabla ^{2}\psi -\sigma \left\vert \psi \right\vert ^{2\nu
}\psi =0,  \label{NLSE}
\end{equation}%
where $\sigma =1$ and $-1$ correspond, respectively, to the self-repulsive
(defocusing) and attractive (focusing) signs of the nonlinearity. Typical
physical realizations of the NLSE feature cubic ($\nu=1$) and quintic ($\nu
=2$) nonlinearities.

The commonly known bright-soliton solutions of the self-focusing ($\sigma
=-1 $) NLSE in 1D, with arbitrary real chemical potential $\mu <0$, are free
of singularities:%
\begin{equation}
\psi =\exp (-i\mu t)\left( \frac{\sqrt{-\left( \nu +1\right) \mu }}{\mathrm{%
\cosh }\left( \sqrt{-2\mu }\nu x\right) }\right) ^{1/\nu }.  \label{soliton}
\end{equation}%
At $\nu \geq 2$, these 1D solutions are subject to instability against the
\textit{wave collapse}, i.e., catastrophic shrinkage of the self-focusing
field \cite{Berge}. In the case of $\sigma =+1$ (self-defocusing), Eq. (\ref%
{NLSE} gives rise to an exact solution in the form of a \textit{singular
soliton},%
\begin{equation}
\psi =\exp \left( -i\mu t\right) \left( \frac{\sqrt{-\left( \nu +1\right)
\mu }}{\mathrm{\sinh }\left( \sqrt{-2\mu }\nu |x|\right) }\right) ^{1/\nu }.
\label{sing-sol}
\end{equation}%
For $\nu \leq 2$, this singular solution is physically irrelevant because it
gives rise to a divergent total norm, $N=\int_{-\infty }^{+\infty
}\left\vert \psi (x)\right\vert ^{2}dx$. In particular, in terms of the
usual GPE, the divergence of the norm implies that the creation of BEC
states in the form of singular solitons (\ref{sing-sol}) would require an
infinite number of atoms. In optics, where the 1D version of Eq. (\ref{NLSE}%
), with $t$ replaced by the propagation distance, $z$, governs paraxial
propagation of a stationary light beam in a planar nonlinear waveguide with
transverse coordinate $x$, the divergence implies an infinite power of the
beam.

On the other hand, at $\nu >2$ the norm of the singular solution (\ref%
{sing-sol}) converges, the exact result being%
\begin{equation}
N_{\sinh }=\frac{\left( \nu +1\right) ^{1/\nu }}{\sqrt{2\pi }\nu }\Gamma
\left( \frac{1}{2}-\frac{1}{\nu }\right) \Gamma \left( \frac{1}{\nu }\right)
\left( -\mu \right) ^{-\left( 1/2-1/\nu \right) },  \label{Nsinh}
\end{equation}%
where $\Gamma $ is the Euler's Gamma-function. Note that, in the limit of $%
\mu \rightarrow -0$, solution (\ref{sing-sol}) takes the form of%
\begin{equation}
\psi _{\mu =0}=\left( \frac{\sqrt{\nu +1}}{\sqrt{2}\nu |x|}\right) ^{1/\nu },
\label{mu=0}
\end{equation}%
whose norm diverges at all values of $\nu ,$ in agreement with Eq. (\ref%
{Nsinh}).

In the $D$-dimensional case, a known mathematical result is that Eq. (\ref%
{NLSE}) with self-repulsion ($\sigma =1$) \cite{Veron} and attraction ($%
\sigma =-1$) \cite{Lions,Gidas}\ gives rise to solutions with singular
density at $r\rightarrow 0$:%
\begin{equation}
|\psi |^{2}\approx \left( \sigma \frac{1+\left( 2-D\right) \nu }{2\nu ^{2}}%
\right) ^{1/\nu }r^{-2/\nu }.  \label{singular}
\end{equation}%
At $D=1$ and $2$, the singular asymptotic form (\ref{singular}) makes sense
for $\sigma =1$ and all values of $\nu $, while it is irrelevant for $\sigma
=-1$. In particular, the singularity of exact solution (\ref{sing-sol})
agrees with Eq. (\ref{singular}), for $\sigma =1$ and $D=1$. The 3D version
of Eq. (\ref{singular}), $|\psi |^{2}\approx \left( \sigma \left( 1-\nu
\right) /\left( 2\nu ^{2}\right) \right) ^{1/\nu }r^{-2/\nu }$, with $\sigma
=1$ and $-1$, admits values, severally, $\nu <1$ and $\nu >1$, and it
vanishes in the fundamentally important case of the cubic nonlinearity, $\nu
=1$. In this case, a more accurate consideration yields \cite{HS}%
\begin{equation}
\left\vert \psi _{\mathrm{3D}}^{\mathrm{(cubic)}}\right\vert ^{2}\approx
\frac{1}{4r^{2}\left\vert \ln \left( r_{0}/r\right) \right\vert },
\label{3D}
\end{equation}%
where $r_{0}$ is an arbitrary radial scale (in terms of the asymptotic
approximation), and appropriate regions are $r\ll r_{0}$ and $r\gg r_{0}$ in
the cases of $\sigma =1$ and $-1$, respectively, i.e., $r_{0}$ should be
chosen as a large or small scale in these two cases. The derivation of
asymptotic expression (\ref{3D}) is briefly presented below, see Eqs. (\ref%
{V}) - (\ref{VVVV}).

The singularity produced by Eq. (\ref{singular}) is physically admissible,
producing a convergent norm, for $\nu >1$ at $D=2$, and for $\nu >2/3$ at $%
D=3$. Thus, in the 3D case only the interval of
\begin{equation}
2/3<\nu \leq 1  \label{2/3}
\end{equation}%
is a physically relevant one for NLSE\ (\ref{NLSE}) with the self-defocusing
nonlinearity. If the norm converges, Eq. (\ref{NLSE}) implies that it is
related to the chemical potential by the following scaling formula:%
\begin{equation}
N\sim \left( -\mu \right) ^{(1/\nu )-(D/2)}.  \label{scaling}
\end{equation}%
It is worthy to note that, at $\nu >2/D$, relation (\ref{scaling}) satisfies
the \textit{anti-Vakhitov-Kolokolov (anti-VK)\ criterion}, $dN/d\mu >0$,
which is a necessary stability condition for a family of bound states
maintained by any self-defocusing nonlinearity \cite{anti} (the VK criterion
proper, $dN/d\mu <0$, is necessary for stability of soliton families created
by self-focusing nonlinearity \cite{VK,Berge}). In particular, the entire
3D\ existence interval (\ref{2/3}) is compatible with the anti-VK condition,
which suggests, in particular, that 3D solutions generated by the cubic
self-repulsion may be stable, which is indeed true (see Section 3 below).

The 3D equation (\ref{NLSE}) with $\nu >1$ and self-focusing sign of the
nonlinearity produces the same scaling relation (\ref{scaling}). However, it
is expected that the respective solution families are unstable, as the
relation \emph{does not} satisfy the VK criterion.

\subsubsection{Solitons pinned to a singular potential}

In the 1D setting, singular solitons may be regularized, leading to
localized states with a finite norm, if a delta-functional attractive
potential with strength $\varepsilon >0$ is added to Eq. (\ref{NLSE}):%
\begin{equation}
i\psi _{t}+(1/2)\psi _{xx}-\left\vert \psi \right\vert ^{2\nu }\psi
=-\varepsilon \delta (x)\psi .  \label{eps}
\end{equation}%
An exact solution to Eq. (\ref{eps}) is%
\begin{equation}
\psi =\exp \left( -i\mu t\right) \left( \frac{\sqrt{-\left( \nu +1\right)
\mu }}{\mathrm{\sinh }\left( \sqrt{-2\mu }\nu \left( |x|+\xi \right) \right)
}\right) ^{1/\nu },  \label{regularized}
\end{equation}%
with shift $\xi $, which removes the singularity in solution (\ref%
{regularized}), defined by relation%
\begin{equation}
\tanh \left( \sqrt{-2\mu }\nu \xi \right) =\sqrt{-2\mu }/\varepsilon .
\label{tanh}
\end{equation}%
The finite amplitude of the regularized solution (\ref{regularized}) is%
\begin{equation}
A\equiv \left\vert \psi (x=0)\right\vert =\left[ (1/2)\left( \nu +1\right)
\left( \varepsilon ^{2}+2\mu \right) \right] ^{1/\left( 2\nu \right) }.
\label{A}
\end{equation}%
As it follows from Eq. (\ref{regularized}), such modes, pinned to the
delta-functional attractive potential, exist in a finite interval of
negative values of the chemical potential:%
\begin{equation}
0<-\mu <\varepsilon ^{2}/2.  \label{interval}
\end{equation}%
Exact solutions for solitons pinned to the same potential, but produced by
Eq. (\ref{eps}) with the opposite (self-focusing) sign in front of the
nonlinear term, were recently considered in Ref. \cite{Zhenya}.

For the exact solution given by Eqs. (\ref{regularized}) and (\ref{tanh}),
it is easy to calculate the norm in the case of the cubic nonlinearity, $\nu
=1$:%
\begin{equation}
N=2\left( \varepsilon -\sqrt{-2\mu }\right) .  \label{Nsimple}
\end{equation}%
Note that the $N(\mu )$ dependence given by Eq. (\ref{Nsimple}) obviously
satisfies the anti-VK criterion.

It is relevant to mention that the attractive delta-functional potential
(generally, a complex one, which includes a local gain) may also produce
stable dissipative solitons (in particular, exact ones) in the framework of
the 1D complex Ginzburg-Landau equation, i.e., NLSE with complex
coefficients in front of the second-derivative and nonlinear terms \cite%
{HK1,HK2,Zezy,JOSAB}.

\subsubsection{New results included in the review}

Recent work \cite{HS} has reported results for families of stable 1D, 2D,
and 3D singular solitons produced by Eq. (\ref{NLSE}) with, respectively,
\textit{septimal} (seventh-order, corresponding to $\nu =3$), quintic, and
cubic self-repulsive nonlinearities. The results, which are directly
relevant to the topic of the present review, are summarized in Section 3. In
the same section we explain that all the relevant nonlinear terms, including
the seemingly ``exotic" septimal one, naturally occur in
physical media (chiefly, in optics). An essential conclusion is similar to
that presented in Section 2 for the singular states pulled to the attractive
center: the repulsive nonlinearity helps to create stable singular GSs with
finite norms. On the other hand, 2D states with embedded vorticity can be
found in the model with the attractive sign of the quintic nonlinearity, but
they are completely unstable.

\section{Two-dimensional singular modes in the attractive potential,
stabilized by the Lee-Huang-Yang (LHY) term}

\subsection{ Analytical approximations}

\subsubsection{The asymptotic form of the solutions at $r\rightarrow 0$ and $%
r\rightarrow \infty $}

Stationary solutions to Eq. (\ref{psi2d}) with integer vorticity $%
l=0,1,2,... $, are looked for as%
\begin{equation}
\psi \left( r,t\right) =\exp \left( -i\mu t+il\theta \right) u(r),
\label{psichi2D}
\end{equation}%
with real radial function obeying the equation%
\begin{equation}
\mu u=-\frac{1}{2}\left( \frac{d^{2}u}{dr^{2}}+\frac{1}{r}\frac{du}{dr}+%
\frac{U_{l}}{r^{2}}u\right) +\sigma u^{3}+u^{4},  \label{chi2D}
\end{equation}%
\begin{equation}
U_{l}\equiv U_{0}-l^{2}.  \label{Ul}
\end{equation}%
At $r\rightarrow 0$, an asymptotic expansion of a relevant solution of Eq. (%
\ref{chi2D}) is
\begin{equation}
u=\left[ \frac{1}{2}\left( U_{l}+\frac{4}{9}\right) \right]
^{1/3}r^{-2/3}-\sigma \frac{9U_{l}+4}{27U_{l}+16}+O\left( r^{2/3}\right) ,
\label{r=0-2D}
\end{equation}%
cf. Eq. (\ref{transition}). Obviously, the density singularity corresponding
to this asymptotic solution, $u^{2}\sim r^{-4/3}$, is weak enough to make
the 2D integral norm (\ref{N}) convergent at $r\rightarrow 0$.

Asymptotic expression (\ref{r=0-2D}) suggests substitution
\begin{equation}
u(r)\equiv r^{-2/3}\chi (r)  \label{uchi}
\end{equation}%
in Eq. (\ref{chi2D}), from which one can derive an equation for the
singularity-free radial functions $\chi (r)$:
\begin{equation}
\mu \chi =-\frac{1}{2}\left[ \frac{d^{2}\chi }{dr^{2}}-\frac{1}{3r}\frac{%
d\chi }{dr}+\frac{\left( U_{l}+4/9\right) }{r^{2}}\chi \right] +\sigma \frac{%
\chi ^{3}}{r^{4/3}}+\frac{\chi ^{4}}{r^{2}}.  \label{chi}
\end{equation}%
Accordingly, the asymptotic form (\ref{chi}) of the solution at $%
r\rightarrow 0$ is replaced by a singularity-free expansion,\textbf{\ }%
\begin{equation}
\chi =\left[ \frac{1}{2}\left( U_{l}+\frac{4}{9}\right) \right]
^{1/3}-\sigma \frac{9U_{l}+4}{27U_{l}+16}r^{2/3}+O\left( r^{4/3}\right) .
\label{r=0-2D_chi}
\end{equation}

Usually, the presence of integer vorticity $l\geq 1$ implies that the
amplitude vanishes at $r\rightarrow 0$ as $r^{l}$, which is necessary
because the phase of the vortex field is not defined at $r=0$. However, the
indefiniteness of the phase is also compatible with the amplitude diverging
at $r\rightarrow 0$. In the linear equation, this is the divergence of the
standard Neumann's cylindrical function, $Y_{l}(r)\sim r^{-l}$, which makes
the respective 2D state unnormalizable for all values $l\geq 1$. In the
present case, Eq. (\ref{r=0-2D}) demonstrates that the interplay of the
attractive potential and quartic self-repulsion curtails the divergence to
the level of $u\sim r^{-2/3}$, for all values of $l$, thus securing the
normalizability of all the states under the consideration. This conclusion
may be compared to what was found in Ref. \cite{HS1}, where the quintic
repulsive term produced another integrable singularity of the 2D density,
with $u(r)\sim r^{-1/2}$.

The asymptotic form of the solution, given by Eq. (\ref{r=0-2D_chi}) is
meaningful if it yields $\chi (r)>0$ [otherwise, Eq. (\ref{chi2D}) cannot be
derived from Eq. (\ref{psi2d})], i.e., for $U_{l}>0$, as well as for $%
-4/9<U_{l}<0$. The latter interval implies, according to Eq. (\ref{Ul}),%
\begin{equation}
l^{2}-4/9<U_{0}<l^{2}.  \label{4/9}
\end{equation}%
For the vortex states, with $l\geq 1$, condition (\ref{4/9}) means, in any
case, $U_{0}>0$. However for the GS with $l=0$, Eq. (\ref{4/9}) admits an
interval of \emph{negative} values of $U_{0}$, namely,
\begin{equation}
0<-U_{0}<4/9.  \label{<4/9}
\end{equation}%
While the existence of the bound state under the combined action of the
repulsive potential, with $U_{0}<0$, and the defocusing quartic nonlinearity
is a counter-intuitive finding, it is closely related to the above-mentioned
fact, first reported in Ref. \cite{Veron}, that the 2D equation with the
self-defocusing nonlinearity acting in the free space (without any
potential) admits singular solutions with asymptotic form (\ref{singular}).
If potential (\ref{U(r)}) is added to the 2D version of Eq. (\ref{NLSE}),
expression (\ref{singular}) is replaced by%
\begin{equation}
\left\vert \psi (r)\right\vert ^{2}\approx \left[ \frac{1}{2}\left( \frac{1}{%
\nu ^{2}}+U_{0}\right) \right] ^{1/\nu }r^{-2/\nu },  \label{p}
\end{equation}%
which is tantamount to Eq. (\ref{r=0-2D}) in the case of $\nu =3/2$.

In the limit of $r\rightarrow \infty $, the asymptotic form of the solution
to Eq. (\ref{chi}) is%
\begin{equation}
\chi (r)\approx \chi _{0}r^{1/6}\exp \left( -\sqrt{-2\mu }r\right) ,
\label{chi0}
\end{equation}%
where $\chi _{0}$ is an arbitrary constant, and $\mu $ must be negative.
Then, a coarse approximation for the global solution can be obtained as an
interpolation bridging asymptotic expressions (\ref{r=0-2D_chi}) and (\ref%
{chi0}):%
\begin{equation}
\chi _{\mathrm{interpol}}(r)=\left[ \frac{1}{2}\left( U_{l}+\frac{4}{9}%
\right) \right] ^{1/3}\exp \left( -\sqrt{-2\mu }r\right) .  \label{inter}
\end{equation}%
As mentioned above, families of localized states are usually characterized
by dependences $N(\mu )$. In particular, calculating norm (\ref{N}) for the
approximate solution given by Eq. (\ref{inter}), one obtains%
\begin{equation}
N_{\mathrm{interpol}}(\mu )=2\pi \int_{0}^{\infty }u_{\mathrm{interpol}%
}^{2}(r)rdr=\pi \Gamma \left( \frac{2}{3}\right) \frac{\left(
U_{l}+4/9\right) ^{2/3}}{\left( -4\mu \right) ^{1/3}},  \label{Ninter}
\end{equation}%
where $\Gamma (2/3)\approx \allowbreak 1.\allowbreak 354$ is the value of
the Gamma-function. Note that this dependence satisfies the anti-VK
criterion.

\subsubsection{The Thomas-Fermi (TF) approximation}

The TF approximation, which, strictly speaking, applies under condition $%
U_{0}\gg 1$ (irrespective of the value of $|\mu |$) amounts to dropping
derivatives in Eq. (\ref{chi}). In fact, the TF approximation may produce
relevant results even when $U_{0}$ is not especially large, see below. It
produces an explicit approximate solution in the case of $\sigma =0$ in Eq. (%
\ref{chi}) (with the nonlinearity represented solely by the LHY term):%
\begin{equation}
\chi _{\mathrm{TF}}(r)=\left\{
\begin{array}{c}
\left[ \left( U_{l}+4/9\right) /2-|\mu |r^{2}\right] ^{1/3},~\mathrm{at}%
~~r<r_{0}\equiv \sqrt{\left( U_{l}+4/9\right) /\left( 2|\mu |\right) }, \\
0,~\mathrm{at}~~r>r_{0}~,%
\end{array}%
\right.  \label{TF}
\end{equation}%
for $\mu <0$. In the limit of $r\rightarrow 0$, Eq. (\ref{TF}) yields the
same exact value, $\chi (r=0)=\left[ \left( U_{l}+4/9\right) /2\right]
^{1/3} $, as given by Eq. (\ref{r=0-2D}). On the other hand, an essential
difference from the full solution is that the TF\ approximation predicts a
finite radius $r_{0}$ of the GS, neglecting the exponentially decaying tail
at $r\rightarrow \infty $, cf. Eq. (\ref{chi0}). Further, the TF
approximation (\ref{TF}) makes it possible to calculate the corresponding $%
N(\mu )$ dependence for the GS family:%
\begin{equation}
N_{\mathrm{TF}}^{(\sigma =0)}(\mu )=2\pi \int_{0}^{r_{0}}\left[ r^{-2/3}\chi
_{\mathrm{TF}}(r)\right] ^{2}rdr=C\frac{U_{l}+4/9}{\left( -\mu \right) ^{1/3}%
},  \label{NTF}
\end{equation}%
where a numerical constant is $C\equiv \pi \int_{0}^{1}\left(
x^{-2}-1\right) ^{2/3}xdx\approx 3.80$, cf. Eq. (\ref{Ninter}).

Note that TF radius $r_{0}$ keeps the same value, as given by Eq. (\ref{TF}%
), in the presence of the MF defocusing cubic term with $\sigma =1$ in Eq. (%
\ref{chi2D}), although the shape of the GS is more complex than one given by
Eq. (\ref{TF}) for $\sigma =0$. In this case, the asymptotic limit of the
respective $N_{\mathrm{TF}}^{(\sigma =1)}(\mu )$ dependence at $\mu
\rightarrow -\infty $ is the same as given by Eq. (\ref{NTF}), while in the
limit of $\mu \rightarrow -0$ it features a weaker singularity:%
\begin{equation}
N_{\mathrm{TF}}^{(\sigma =1)}(\mu )\approx \left( \pi /2\right) U_{l}\ln
\left( 1/|\mu |\right) .
\end{equation}

Even in the case of the focusing sign of the MF term, corresponding to $%
\sigma =-1$ in Eq. (\ref{chi2D}), the LHY-induced quartic nonlinearity is
able to stabilize the condensate against the combined action of the MF
self-attraction and pull to the center. In this case, the TF approximation,
applied to Eq. (\ref{chi2D}), cannot be easily resolved to predict $u_{%
\mathrm{TF}}(r)$, but it readily produces an inverse dependence, for $r$ as
a function of $u$:%
\begin{equation}
r^{2}=\left( U_{l}/2\right) \left( -\mu -u^{2}+u^{3}\right) ^{-1}.  \label{r}
\end{equation}%
Then, looking for a maximum of expression (\ref{r}), which is attained at $%
u_{\max }=2/3$, it is easy to find the corresponding size of the TF state:
\begin{equation}
r_{0}^{(\sigma =-1)}=\frac{U_{l}}{2\left( |\mu |-4/27\right) }.  \label{-1}
\end{equation}%
Equation (\ref{-1}) suggests that the GS exists,\ in the case of $\sigma =-1$%
, for $|\mu |$ exceeding a threshold value,
\begin{equation}
|\mu |>\left( |\mu |\right) _{\mathrm{thr}}=4/27.  \label{thr}
\end{equation}%
According to Eq. (\ref{-1}), the norm diverges at $\mu \rightarrow -4/27$ as%
\begin{equation}
N\approx 2\pi \left( r_{0}^{(\sigma =-1)}\right) ^{2}u_{\max }^{2}=\frac{%
2\pi }{9}\frac{U_{l}^{2}}{\left( |\mu |-4/27\right) ^{2}}.  \label{NN}
\end{equation}

The analytical predictions reported in this subsection are compared to their
numerical counterparts in the following one.

\subsection{Numerical results for the 2D modes stabilized by the LHY term}

Stationary solutions of Eq. (\ref{chi}) could be readily produced by means
of the Newton's iteration method. Then, stability of stationary solutions
was identified by solution of the linearized eigenvalue problem for small
perturbations, represented by terms
\begin{equation}
\delta \psi \sim \exp \left( -i\mu t+\Lambda t\right) ,~\exp \left( -i\mu
t+\Lambda ^{\ast }t\right)  \label{pert}
\end{equation}%
added to the stationary states ($\ast $ stands for the complex conjugation),
the stability condition being, as usual \cite{Yang}, that all eigenvalues $%
\Lambda $ must be pure-imaginary. The so predicted (in)stability was
verified by direct simulations of Eq. (\ref{psi2d}). The results were
produced for the model including the cubic term in Eq. (\ref{psi2d}), with $%
\sigma =\pm 1$, as well as for the most fundamental case of $\sigma =0$,
when the nonlinearity is represented solely by the LHY term.

\begin{figure}[tbp]
\subfigure[]{\includegraphics[width=3.2in]{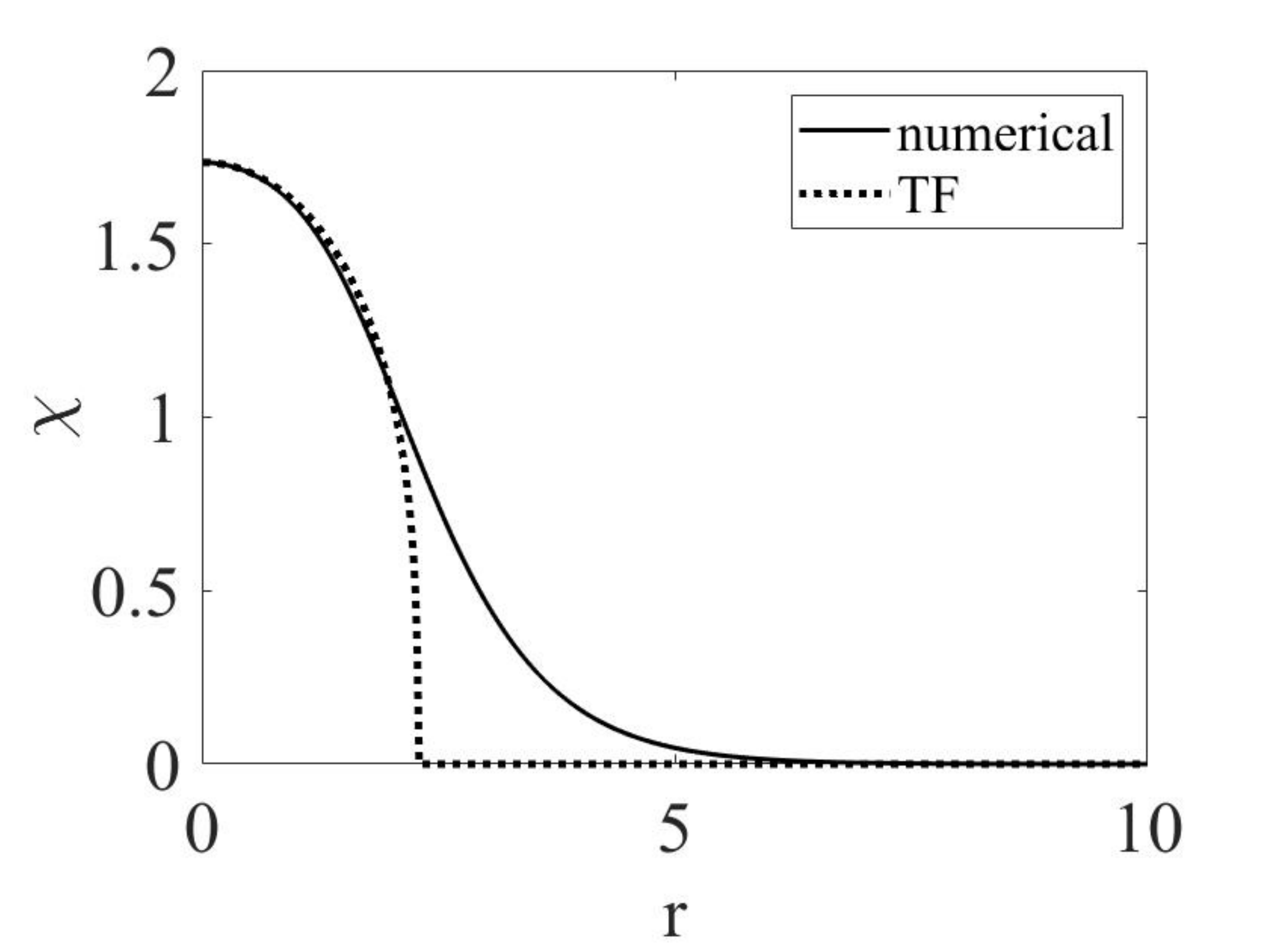}}%
\subfigure[]{
\includegraphics[width=3.2in]{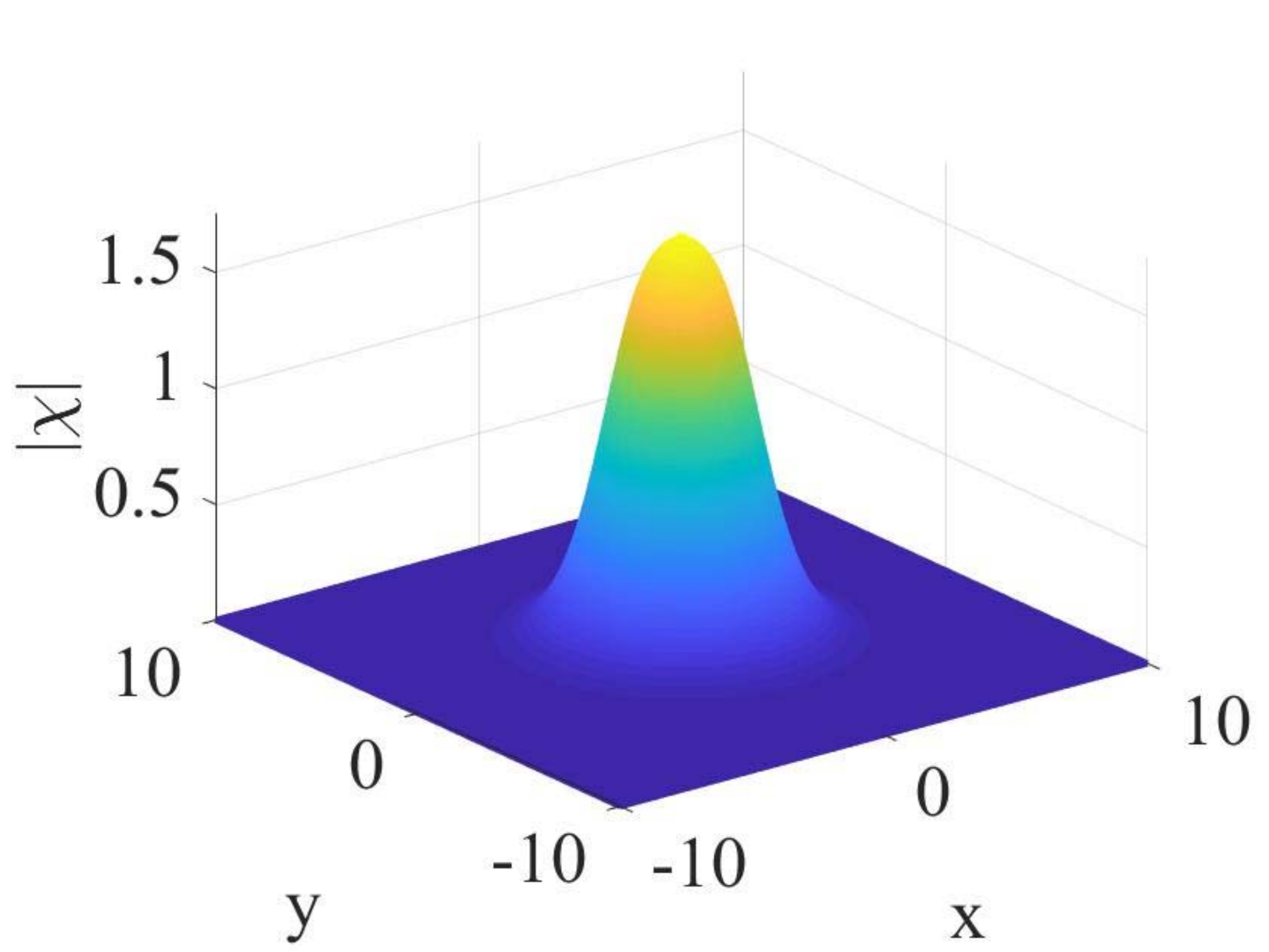}} 
\caption{(a) The profile of a (stable) numerically found GS and its TF
counterpart, produced by Eqs. (\protect\ref{chi}) and (\protect\ref{TF}),
respectively, for $\protect\sigma =0$, $U_{0}=10$, and $\protect\mu =-1$.
The norms of the numerical and approximate solutions are $N_{\mathrm{num}%
}=41.05$, $N_{\mathrm{TF}}=39.68$. (b) The global view of the numerical
solution.}
\label{fig1a}
\end{figure}

A typical stable GS, obtained as a numerical solution of Eq. (\ref{chi})
with $\sigma =0$, $U_{0}=10$, and $\mu =-1$, is displayed in Fig. \ref{fig1a}%
, together with its counterpart produced by the TF approximation (the
interpolation-based approximation, given by Eq. (\ref{inter}), is not
displayed here, as it is not relevant for large values of $U_{0}$).\ It is
seen that the TF approximation is very close to its numerical counterpart in
the inner zone, $r<r_{0}$ (see Eq. (\ref{TF})), while in the outer one, $%
r>r_{0}$, the TF approximation yields zero, being inaccurate, in this sense.
Actually, the overall impact of the difference between the numerical and TF
solutions in the outer zone is less significant due to the presence of
factor $r^{-2/3}$ in expression (\ref{uchi}) for the full solution, $u(r)$.
As a result, in the case presented in Fig. \ref{fig1a} the relative error of
the norm of the TF\ solution, calculated as per Eq. (\ref{NTF}), is $\approx
0.03$ .

\begin{figure}[tbp]
{\includegraphics[width=3.2in]{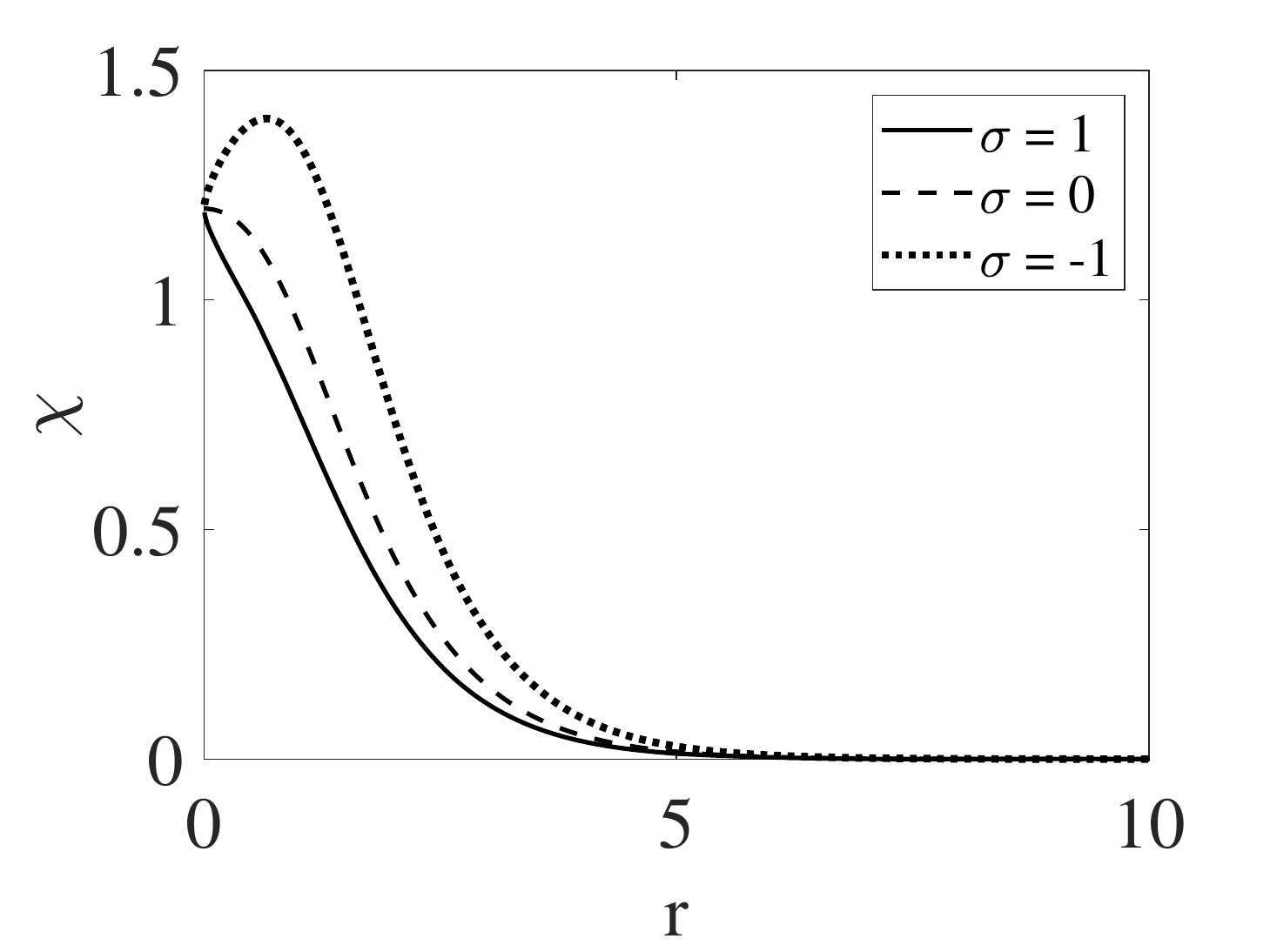}}
\caption{Numerically generated stable GS solutions of Eq. (\protect\ref{chi}%
) for $U_{0}=3,\protect\mu =-0.8$ , and $\protect\sigma =1$, $0$, and $-1$.
The respective norms are $N(\protect\sigma =1)=11.7$, $N(\protect\sigma %
=0)=15.41$ and $N(\protect\sigma =-1)=24.62$.}
\label{fig2a}
\end{figure}

The effect of the MF cubic term of either sign, repulsive ($\sigma =1$) or
attractive ($\sigma =-1$), in comparison with the case of $\sigma =0$, on
the GS is shown in Fig. \ref{fig2a}. At $r=0$, all the three shapes converge
to a common value, $\chi (r=0)\approx 1.20$, in agreement with Eq. (\ref%
{r=0-2D_chi}).

\begin{figure}[tbp]
\subfigure[]{\includegraphics[width=3.2in]{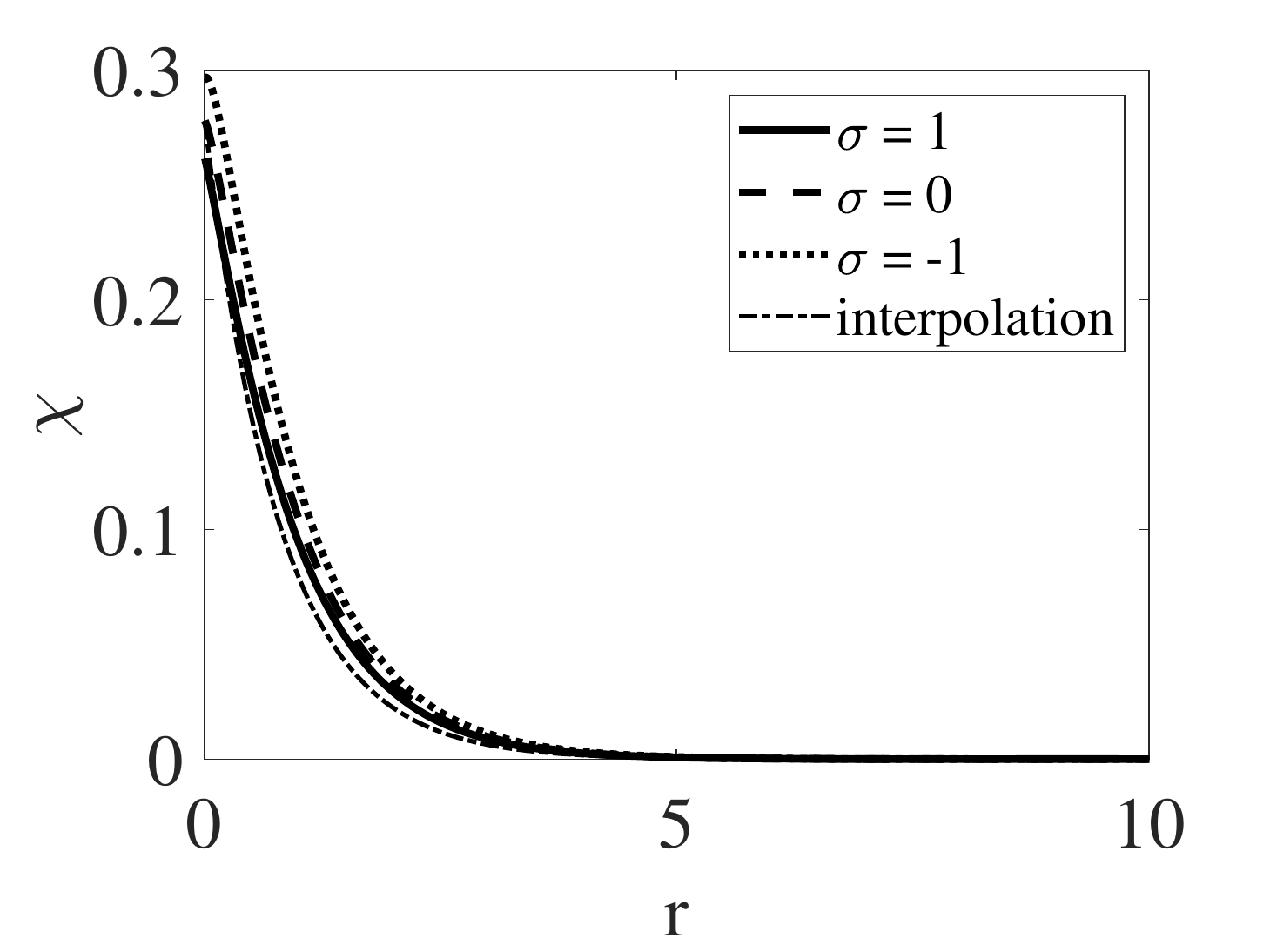}}\subfigure[]{%
\includegraphics[width=3.2in]{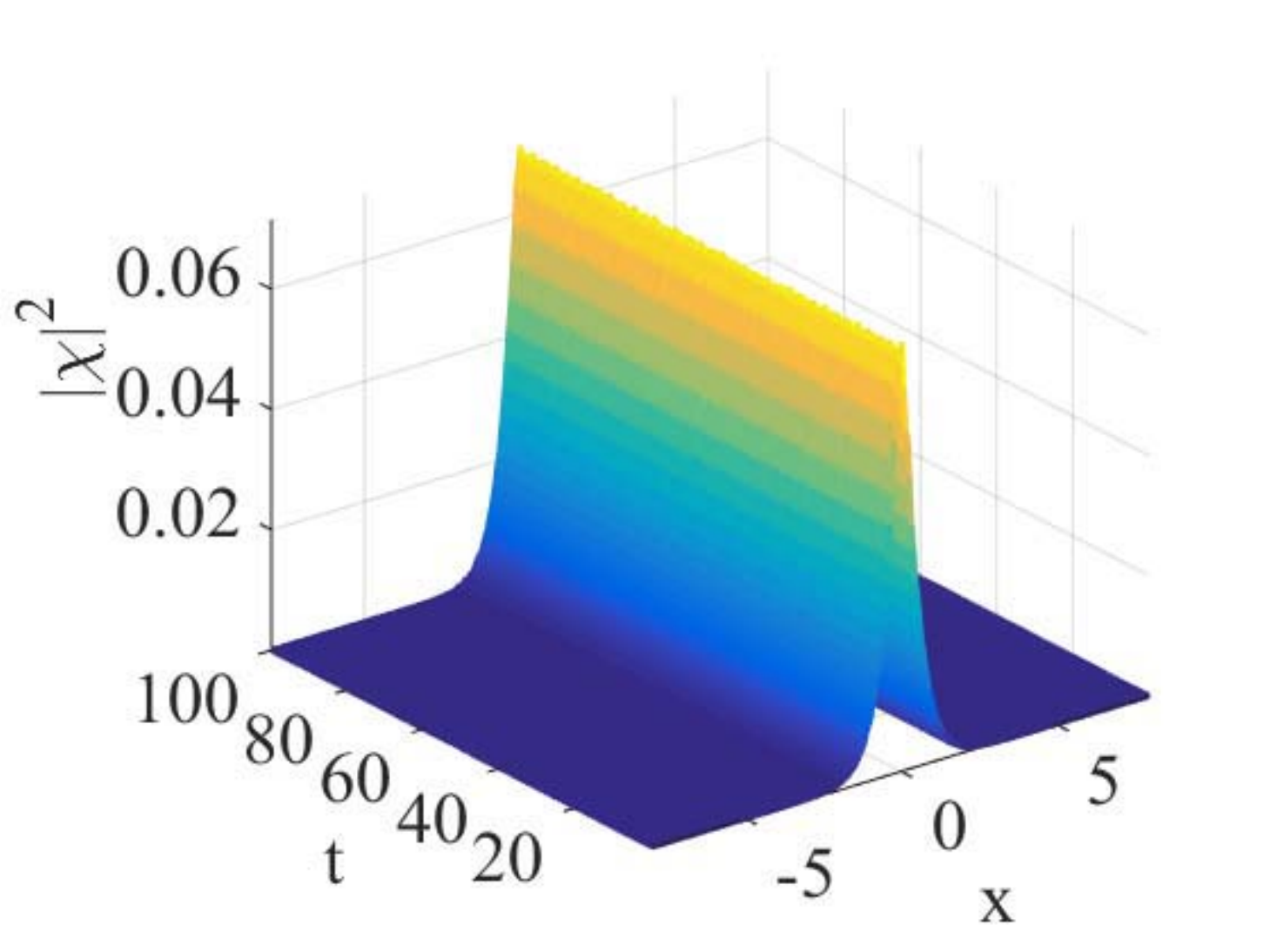}}
\caption{(a) Radial profiles of numerically found GSs, produced by Eq. (%
\protect\ref{chi}), and the respective interpolation profile (\protect\ref%
{inter}), in the presence of the \emph{repulsive potential }in Eq. (\protect
\ref{chi}), with $U_{0}=-0.40$, at $\protect\mu =-0.8$, without and with the
repulsive or attractive MF cubic term ($\protect\sigma =0$ and $\protect%
\sigma =1$ or $-1$, respectively). The corresponding values of the norm are $%
N\left( \protect\sigma =1\right) =0.41$, $N(\protect\sigma =0)=0.45$, and $N(%
\protect\sigma =-1)=0.52$. The interpolating approximation gives $N\approx
0.36$, as per Eq. (\protect\ref{Ninter}). (b) Stability of the GS mode in
direct simulations, in the case of $\protect\sigma =1$.}
\label{fig3a}
\end{figure}

The counter-intuitive prediction of the existence of the GS in the presence
of the \emph{repulsive} potential in Eq. (\ref{chi}), with $U_{0}$ belonging
to interval (\ref{<4/9}), was also confirmed by the numerical results.
Figure \ref{fig3a} displays numerically found GSs for $U_{0}=-0.40$, taken
close to the limit value, $U_{0}=-4/9\approx \allowbreak -0.44$, for all the
three values of the coefficient in front of the MF cubic term in Eq. (\ref%
{chi}), \textit{viz}., $\sigma =0$ and $\pm 1$. The same figure shows that
the interpolating approximation for these solutions, provided by Eq. (\ref%
{inter}), is quite accurate in this case (the TF approximation (\ref{TF})
does not apply to $U_{0}<0$). It is seen too that the MF cubic term does not
strongly affect the solution.

Dependences $N(\mu )$ for families of the GS solutions, obtained from Eq. (%
\ref{chi}) without and with the repulsive or attractive MF cubic term ($%
\sigma =0$ and $\sigma =\pm 1$, respectively), are displayed in Figs. \ref%
{fig4a}(a,b), for different values of strength $U_{0}$ of the potential,
both $U_{0}>0$ and $U_{0}<0$. In panel (a), the $N(\mu )$ curves produced by
the TF approximation as per Eq. (\ref{NTF}) are compared to their numerical
counterparts. The same panel demonstrates that the interpolating
approximation is very accurate for $U_{0}=-0.4$ (while its accuracy is poor
for $U_{0}>0$).

\begin{figure}[tbp]
\subfigure[]{\includegraphics[width=3.2in]{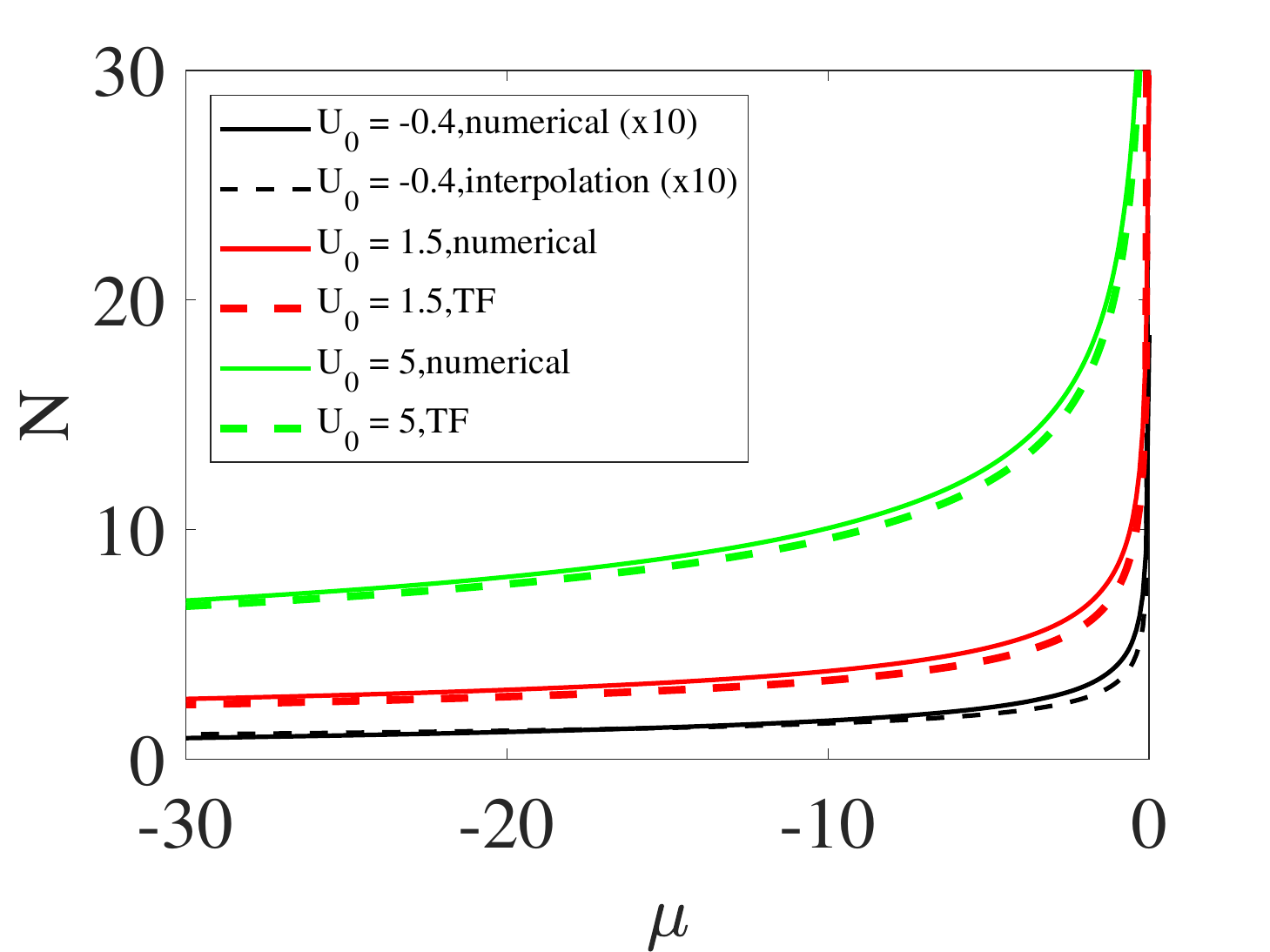}} \subfigure[]{%
\includegraphics[width=3.2in]{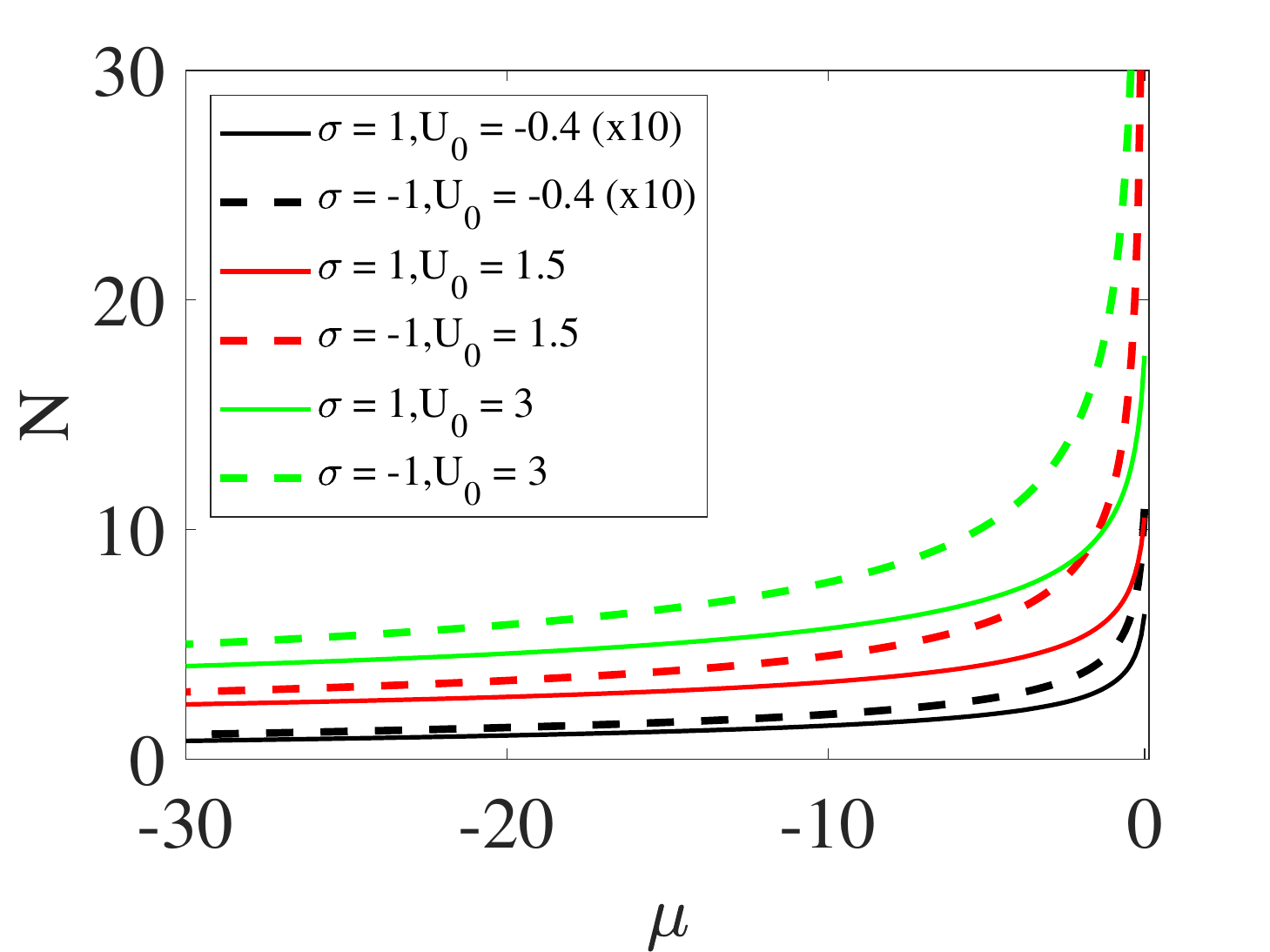}} \subfigure[]{%
\includegraphics[width=3.2in]{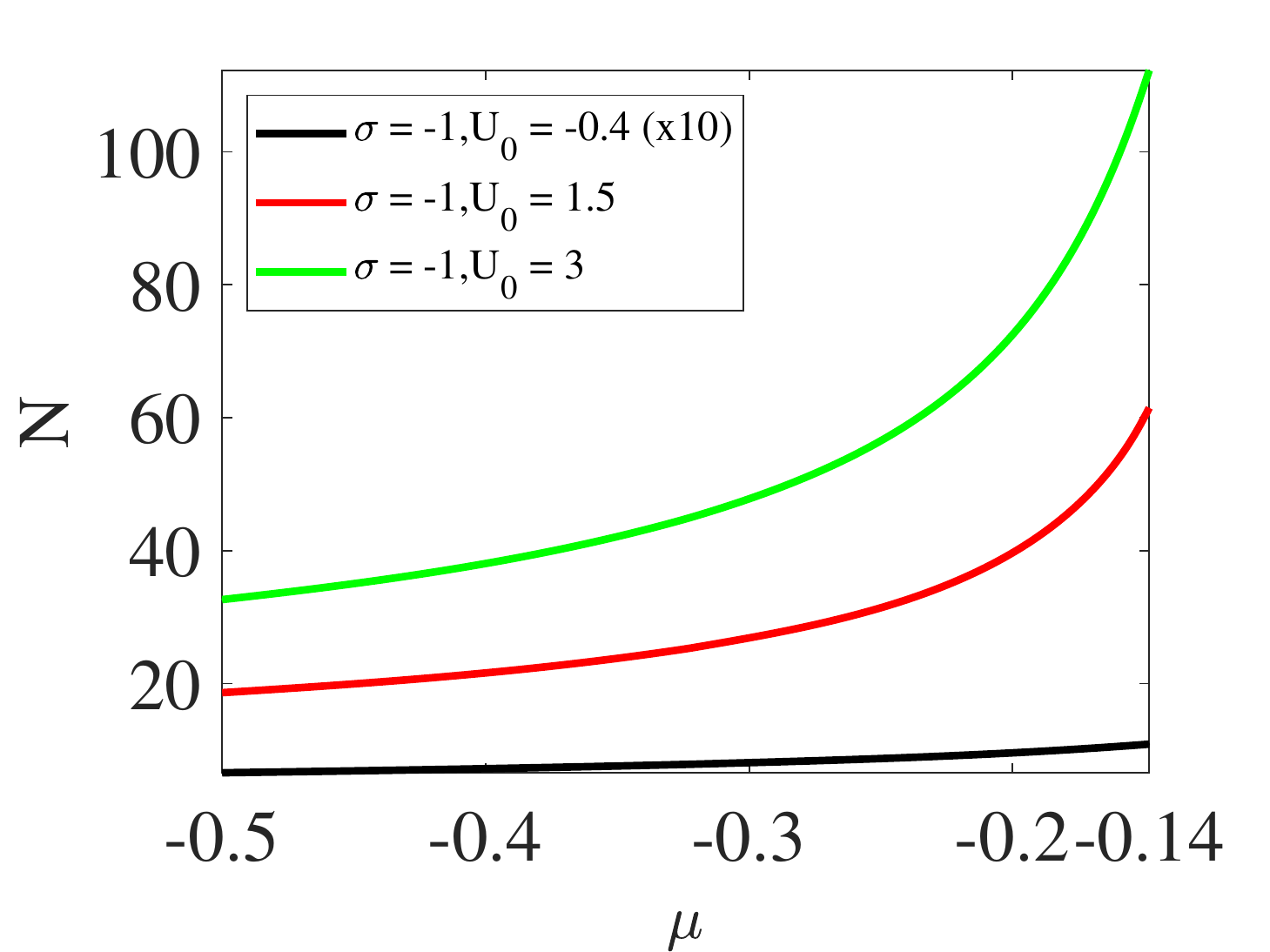}}
\caption{Dependences $10\times N(\protect\mu )$ for stable GS solutions with
$U_{0}=-0.4$, and $N(\protect\mu )$ for $U_{0}=$ $1.5$, $3.0$, $5.0$, which
correspond, respectively, to the repulsive and attractive central potential
(in the former case, $N$ is multiplied by $10$, as the actual values of the
norm are too small in this case). Panels (a) and (b) correspond to the
system which does not or does include the repulsive or attractive cubic term
($\protect\sigma =0$ and $1,-1$, respectively). In (a), the numerical
results are juxtaposed with the TF counterparts produced by Eq. (\protect\ref%
{NTF}) (except for $U_{0}=-0.4$, when the TF approximation is irrelevant;
however, in this case the interpolating approximation, based on Eq. (\protect
\ref{Ninter}), is very close to the numerically generated curve). In (b),
the numerical curves are compared for $\protect\sigma =1$ and $-1$. Panel
(c) is a zoom of the plot from (b) at small values of $|\protect\mu |$, with
the aim to show relative proximity to the threshold value $\left( |\protect%
\mu |\right) _{\mathrm{thr}}=4/27\approx \allowbreak 0.15$ for $\protect%
\sigma =-1$, as predicted in the TF approximation by Eq. (\protect\ref{thr})
(this point is discussed in detail in the text).}
\label{fig4a}
\end{figure}

Panel (c) in Fig. \ref{fig4a} confirms that, in the presence of the
attractive cubic term ($\sigma =-1$), the GS exists at $|\mu |>\left( |\mu
|\right) _{\mathrm{thr}}$, as predicted by the TF approximation in Eq. (\ref%
{thr}). Up to the accuracy of the numerical results, the threshold value is
indeed $\left( |\mu |\right) _{\mathrm{thr}}=4/27$, as given by Eq. (\ref%
{thr}). This finding is explained by the fact that the width of the GS
diverges in the limit of $|\mu |\rightarrow \left( |\mu |\right) _{\mathrm{%
thr}}$, as seen in Eq. (\ref{-1}), hence in this limit the derivatives
become negligible in Eq. (\ref{chi2D}), making the TF approximation
asymptotically exact. In principle, $N(\mu )$ must steeply diverge at $|\mu
|\rightarrow 4/27$, according to Eq. (\ref{NN}), but it is difficult to
collect numerical data very close to the threshold, as the GS is extremely
broad in this limit.

The computation of eigenvalues for small perturbations, as well as direct
simulations, demonstrate full stability of the GS solutions for both
positive and negative values of $U_{0}$ (at which the GS exists), at all $%
\mu <0$. In particular, Fig. \ref{fig3a}(b) demonstrates the stability in
the counter-intuitive case of the repulsive potential, with $U_{0}=-0.4$.
The stability is not affected either by the TF cubic term, holding for $%
\sigma =0$ and $\pm 1$.

Lastly, an analytical consideration, verified by numerical data \cite{we}, has demonstrated that
vortex modes with $l=1$ (see Eq. (\ref{psichi2D})), are stable at $U_0>14/9$,
and unstable at $U_0<14/9$. For the vortices with $l=2$, the critical value
of the pulling-potential strength, below which they are unstable, is much larger,
\textit{viz}., $U_0=77/9$.  Simulations of the evolution of unstable vortex
modes demonstrate that the vortical pivot drifts in the outside direction, along
an unwinding spiral trajectory. Eventually, the pivot is ousted to
periphery, and the vortex mode transforms into a stable GS with zero
vorticity. In this case, the norm of the residual GS is essentially smaller than the
original value, due to intensive emission of small-amplitude waves in the
course of the transient evolution.

\section{Singular solitons in one, two, and three dimensions}

This section summarizes original results which were recently reported in
Ref. \cite{HS}. In the 2D case, the results are related to those presented
in the previous section for the model with the external potential.

As mentioned in the Introduction, relevant self-defocusing nonlinearities
which are necessary to support singular self-trapped states in 3D, 2D, and
1D settings, are represented by the cubic, quintic, or septimal terms, that
correspond, respectively, to $\nu =1$, $2$, and $3$ in Eq. (\ref{NLSE}).
Realizations of the cubic and quintic nonlinearities of either sign
(focusing or defocusing) in diverse physical media are well known \cite%
{Michinel,Anger,review,Brazil}. In particular, such nonlinear terms with
controllable strengths (including the ``exotic" septimal
term), can be realized in optical waveguides filled by suspensions of
metallic nanoparticles, control parameters being the density and size of the
particles \cite{Cid2,Cid1,Cid3}.

Results outlined below provide not only solutions in analytical and
numerical forms, but also an interpretation of the physical meaning of the
singular solitons.

\subsection{Analytical results}

\subsubsection{The one-dimensional model with the septimal nonlinearity}

Singular solitons created by the 1D version of Eq. (\ref{NLSE}) with the
seventh-order defocusing ($\sigma =1$) nonlinearity, $\nu =3$, are looked
for as
\begin{equation}
\psi \left( x,t\right) =\exp \left( -i\mu t\right) \mathrm{U}(x),  \label{uU}
\end{equation}%
with real function $\mathrm{U}(x)$ satisfying equation%
\begin{equation}
\frac{1}{2}\frac{d^{2}\mathrm{U}}{dx^{2}}=-\mu \mathrm{U}+\mathrm{U}^{7}.
\label{U}
\end{equation}%
In the application to the planar optical waveguide, $t$ is not time, but the
propagation distance (often denoted $z$), $x$ is the transverse coordinate,
and $-\mu $ is the propagation constant.

The exact solution of Eq. (\ref{U}) is given by Eq. (\ref{sing-sol}):%
\begin{equation}
\mathrm{U}(x)=\left( \frac{2\sqrt{-\mu }}{\mathrm{\sinh }\left( 3\sqrt{-2\mu
}|x|\right) }\right) ^{1/3}.  \label{Uexact}
\end{equation}%
The asymptotic form of solutions to Eq. (\ref{U}) at $x\rightarrow 0$ does
not depend on $\mu $,

\begin{equation}
\mathrm{U}(x)\approx \left( 2/9\right) ^{1/6}|x|^{-1/3}.  \label{-1/3}
\end{equation}%
Note that expression (\ref{-1/3}) is an exact solution of Eq. (\ref{U}) with
$\mu =0$, but its integral norm diverges at $|x|\rightarrow \infty $. For $%
\mu <0$, the exponentially decaying asymptotic form of Eq. (\ref{Uexact}) at
$|x|\rightarrow \infty $ is
\begin{equation}
\mathrm{U}(x)\approx \left( 4\sqrt{-\mu }\right) ^{1/3}\exp \left( -\sqrt{%
-2\mu }|x|\right) .  \label{tail}
\end{equation}

Lastly, the norm of the 1D soliton family is given by Eq. (\ref{Nsinh}),%
\begin{equation}
N=\frac{2^{2/3}}{3\sqrt{2\pi }}\Gamma \left( \frac{1}{6}\right) \Gamma
\left( \frac{1}{3}\right) \left( -\mu \right) ^{-1/6}\approx \allowbreak
3.1478\left( -\mu \right) ^{-1/6},  \label{N1D}
\end{equation}%
with the numerical coefficient which is accidentally close to $\pi $. This $%
N(\mu )$ dependence satisfies the anti-VK criterion, which, as mentioned
above, is necessary for the stability of localized modes supported by
repulsive nonlinearities \cite{anti}.

\subsubsection{Physical interpretation of the 1D singular soliton: screening
of a ``bare" $\protect\delta $-functional potential}

Although the existence of the stable singular solitons under the action of
the septimal self-repulsive nonlinearity is firmly established by the above
analysis, this result may seem counter-intuitive, as it is commonly believed
that localized modes may only be supported by self-attraction. The purport
of the result may be understood by comparing Eq. (\ref{U}) to a modified
equation,%
\begin{equation}
\frac{1}{2}\frac{d^{2}\mathrm{U}}{dx^{2}}=-\mu \mathrm{U}+\mathrm{U}%
^{7}-\varepsilon \delta (x)\mathrm{U},  \label{eps2}
\end{equation}%
which includes an attractive delta-functional potential with strength $%
\varepsilon >0$, cf. Eq. (\ref{eps}). An exact solution to Eq. (\ref{eps2})
is produced by Eqs. (\ref{regularized}), (\ref{tanh}), and (\ref{A}):%
\begin{equation}
\mathrm{U}(x)=\left( \frac{2\sqrt{-\mu }}{\mathrm{\sinh }\left( 3\sqrt{-2\mu
}\left( |x|+\xi \right) \right) }\right) ^{1/3},  \label{U nu=3}
\end{equation}%
\begin{equation}
\xi =\frac{1}{6\sqrt{-2\mu }}\ln \left( \frac{\varepsilon +\sqrt{-2\mu }}{%
\varepsilon -\sqrt{-2\mu }}\right) \approx \frac{1}{3\varepsilon }.
\label{xi nu=3}
\end{equation}%
\begin{equation}
A=\left[ 2\left( \varepsilon ^{2}+2\mu \right) \right] ^{1/6}  \label{A nu=3}
\end{equation}%
(recall $A\equiv \left\vert \mathrm{U}(x=0)\right\vert $ is the amplitude of
the mode pinned to the attractive delta-functional potential). The
approximate value of the offset in Eq. (\ref{xi nu=3}) corresponds to the
limit of $\varepsilon \gg \sqrt{-2\mu }$.

This solution may be considered as a regularized version of its singular
counterpart (\ref{-1/3}), which obviously converges to the singular state
for $\varepsilon \rightarrow \infty $. It is also relevant to calculate the
value of the Hamiltonian, corresponding to Eq. (\ref{eps2}),%
\begin{equation}
H_{\varepsilon }=\int_{-\infty }^{+\infty }\left[ \frac{1}{2}\left( \frac{d%
\mathrm{U}}{dx}\right) ^{2}+\frac{1}{4}|\mathrm{U}|^{8}\right]
dx-\varepsilon |\mathrm{U}(x=0)|^{2}.  \label{H}
\end{equation}%
In the limit of $\varepsilon \gg \sqrt{-2\mu }$, it is
\begin{equation}
H_{\varepsilon }\approx -\left( 1/5\right) \left( 3\varepsilon \right)
^{5/3},  \label{Heps}
\end{equation}%
the negative sign suggesting that solution may be the GS.

A physical interpretation of the ability of the self-repulsive 1D model to
create singular solitons is offered by the fact that, according to Eq. (\ref%
{xi nu=3}), the strength (``bare charge") $\varepsilon $ of
the delta-functional attractive potential diverges in the limit of $\xi
\rightarrow 0$, which brings one back to the underlying singular solution
given by Eqs. (\ref{-1/3})-(\ref{tail}). This observation implies that an
\emph{infinitely large} ``bare attractive charge", embedded
in the self-defocusing septimal medium, is \emph{completely screened} by the
nonlinearity, which builds the singular soliton with the convergent norm,
for that purpose. This mechanism roughly resembles the renormalization
procedure in quantum electrodynamics, where an infinite \textit{bare charge}
of the electron cancels with other diverging factors, making it possible to
produce finite observable predictions.

\subsubsection{The two-dimensional model with the quintic nonlinearity}

In the 2D setting, it is relevant to consider Eq. (\ref{NLSE}) with the
quintic self-defocusing term:
\begin{equation}
i\psi _{t}=-\frac{1}{2}\nabla ^{2}\psi +|\psi |^{4}\psi .  \label{u2D}
\end{equation}%
In terms of optics, Eq. (\ref{u2D}) models the paraxial propagation of light
in a bulk waveguide, with $t$ being the propagation distance (usually
denoted $z$).

In polar coordinates $\left( r,\theta \right) $, solutions of Eq. (\ref{u2D}%
) with integer vorticity, $l=0,1,2,...$, is looked for as
\begin{equation}
\psi =\exp \left( -i\mu t+il\theta \right) \mathrm{U}(r),  \label{M}
\end{equation}%
with real amplitude function $\mathrm{U}(r)$ satisfying the radial equation,
cf. Eq. (\ref{U}):%
\begin{equation}
\frac{1}{2}\left( \frac{d^{2}\mathrm{U}}{dr^{2}}+\frac{1}{r}\frac{d\mathrm{U}%
}{dr}-\frac{l^{2}}{r^{2}}\mathrm{U}\right) =-\mu \mathrm{U}+\mathrm{U}^{5}.
\label{U2D}
\end{equation}%
For 2D singular solitons with $l=0$, Eq. (\ref{U2D}) produces the following
expansion at $r\rightarrow 0$:
\begin{equation}
\mathrm{U}_{\mathrm{2D}}(r)\approx 2^{-3/4}r^{-1/2}-2^{1/4}\mu r^{3/2},
\label{2Drep}
\end{equation}%
with the respective 2D integral norm converging at $r\rightarrow 0$. For $%
\mu =0$,
\begin{equation}
\mathrm{U}_{\mathrm{2D}}^{(\mu =0)}(r)=2^{-3/4}r^{-1/2}  \label{u0}
\end{equation}%
is an exact solution of Eq.~(\ref{U2D}), but its norm diverges at $%
r\rightarrow \infty $. The asymptotic form of the solution at $r\rightarrow
\infty $ is found from the linearized version of Eq. (\ref{U2D}),%
\begin{equation}
\mathrm{U}_{\mathrm{2D}}(r)\approx \frac{C}{\sqrt{r}}\left( 1-\frac{1}{8%
\sqrt{-2\mu }r}\right) \exp \left( -\sqrt{-2\mu }r\right)  \label{2Dexp}
\end{equation}%
[cf. Eq. (\ref{tail})] where $C$ is a constant, and the second term in the
parenthesis its a correction to the lowest approximation.

For vortex states given by Eq. (\ref{M}) with $l\geq 1$, a singular solution
with convergent norm is obtained with the \emph{opposite} (self-focusing)
sign in front of the quintic term in Eq. (\ref{u2D}), the asymptotic
approximation at $r\rightarrow 0$ being
\begin{equation}
\mathrm{U}_{\mathrm{2D}}^{(l)}(r)\approx \left[ \frac{1}{2}\left( l^{2}-%
\frac{1}{4}\right) \right] ^{1/4}r^{-1/2},  \label{2Dattr}
\end{equation}%
cf. Eq. (\ref{2Drep}). For $\mu =0$, Eq. (\ref{2Dattr}) represents an exact
solution of Eq. (\ref{U2D}) with the opposite (self-focusing) sign in front
of the quintic term, but its integral norm diverges at $r\rightarrow \infty $%
.

An exact scaling relation for the norm of the solutions, with $l=0$ and $%
l\geq 1$ alike, follows from Eq. (\ref{u2D}) with either sign of the quintic
term:
\begin{equation}
N_{\mathrm{2D}}(\mu )=\mathrm{const}/\sqrt{-\mu }  \label{N2D}
\end{equation}%
(for $l=0$, a numerically found value is $\mathrm{const}\approx 21.2$). This
dependence satisfies the anti-VK criterion, $dN/d\mu >0$ hence the
respective GS\ solutions with $l=0$, maintained by the defocusing quintic
nonlinearity, may be (and indeed are) stable, but it contradicts the VK
condition per se, $dN/d\mu <0$, hence the vortex states, which exist in the
case of self-focusing, are definitely unstable, as corroborated by numerical
simulations \cite{HS}.

\subsubsection{Interpretation of the 2D singular soliton: screening of a
ring-shaped attractive potential}

Similar to what is outlined above for the 1D model, it is possible to
introduce a version of Eq. (\ref{u2D}) with a delta-functional potential;
however, it is concentrated on a ring of a small radius, $\rho $, instead of
the single point. The so modified equation (\ref{U2D})\ with $l=0$ is%
\begin{equation}
\frac{1}{2}\left( \frac{d^{2}\mathrm{U}}{dr^{2}}+\frac{1}{r}\frac{d\mathrm{U}%
}{dr}\right) =-\mu \mathrm{U}+\mathrm{U}^{5}-\varepsilon _{\mathrm{2D}%
}\delta (r-\rho )\mathrm{U}=0,  \label{2Deps}
\end{equation}%
cf. Eq. (\ref{eps2}). At $r>\rho $ the solution of Eq. (\ref{2Deps}) is
sought for in a form similar to that given by Eq. (\ref{2Drep}), while
inside the ring it is taken as $\mathrm{U}=\mathrm{const}\approx
2^{-3/4}\rho ^{-1/2}$. Eventually, straightforward manipulations, which take
into regard the jump of $d\mathrm{U}/dr$ at $r=\rho $, yield a relation
between the strength of the delta-functional potential concentrated on the
ring and the ring's radius:
\begin{equation}
\varepsilon _{\mathrm{2D}}=1/\left( 4\rho \right) ,  \label{eps2D}
\end{equation}%
cf. Eq. (\ref{xi nu=3}). Then, in the limit of $\rho \rightarrow 0$, an
effective ``charge" of the 2D attractive potential is%
\begin{equation}
Q_{\mathrm{2D}}=2\pi \rho \cdot \varepsilon _{\mathrm{2D}}=\frac{\pi }{2}.
\label{Q2D}
\end{equation}%
Thus, the 2D singular soliton may be construed as a solution providing the
screening of the finite ``charge" by the defocusing quintic
nonlinearity.

\subsubsection{Effects of additional nonlinear terms on 1D and 2D singular
solitons}

Even if the septimal or quintic term represents the dominant nonlinearity in
the underlying NLSE, the lower-order terms, i.e., cubic and/or quintic ones
(with respective coefficients $g_{3}$ and $g_{5}$), should be included in
the realistic model of the light propagation \cite{Cid2,Cid1,Cid3}. In
particular, the accordingly amended septimal 1D equation (\ref{U}) is
replaced by
\begin{equation}
\frac{1}{2}\frac{d^{2}\mathrm{U}}{dx^{2}}=-\mu \mathrm{U}+g_{3}\mathrm{U}%
^{3}+g_{5}\mathrm{U}^{5}+\mathrm{U}^{7}.  \label{1D amended}
\end{equation}%
The additional terms produce negligible corrections to the asymptotic
singular form (\ref{-1/3}) at $x\rightarrow 0$ \cite{HS}:%
\begin{equation}
\mathrm{\delta U}_{\mathrm{1D}}(x)\approx -\left( \frac{32}{81}\right) ^{1/6}%
\frac{g_{5}}{5}|x|^{1/3}-\frac{3g_{3}}{7\sqrt{2}}|x|.  \label{delta1D}
\end{equation}%
Further, if, in the 2D setting, the cubic term (the same as in Eq. (\ref{1D
amended})) is added to Eq. (\ref{U2D}), it also produces a negligible
correction to the singular asymptotic form given by Eq. (\ref{2Drep}) \cite%
{HS}: $\mathrm{\delta U}_{\mathrm{2D}}(r)\approx -2^{-5/4}g_{3}r^{1/2}$, cf.
Eq. (\ref{delta1D}).

\subsubsection{The 3D model with the cubic nonlinearity}

In the 3D case, the relevant equation is the NLSE with the usual cubic term:
\begin{equation}
i\psi _{t}=-\frac{1}{2}\nabla ^{2}\psi +\sigma |\psi |^{2}\psi .  \label{u3D}
\end{equation}%
It has a plethora of physical realizations for both the defocusing ($\sigma
=1$) and focusing ($\sigma =-1$) signs of the nonlinearity \cite{review},
including the GPE for BEC with, respectively, repulsive or attractive
interactions between atoms \cite{GP}. For isotropic stationary states, $\psi
=\exp (-i\mu t)\mathrm{U}(r)$, where $r$ is the 3D radial coordinate, the
real radial functions obeys the commonly known equation,%
\begin{equation}
\frac{1}{2}\left( \frac{d^{2}\mathrm{U}}{dr^{2}}+\frac{2}{r}\frac{d\mathrm{U}%
}{dr}\right) -\sigma \mathrm{U}^{3}=-\mu \mathrm{U}.  \label{radial}
\end{equation}%
For the asymptotic consideration of singular solutions at $r\rightarrow 0$,
the term on the right-hand side of Eq. (\ref{radial}) is negligible.
Dropping this term and looking for solutions in the form of%
\begin{equation}
\mathrm{U}(r)=r^{-1}\mathrm{V}\left( \tau \equiv -\ln (r/r_{0})\right) ,
\label{V}
\end{equation}%
where $r_{0}$ is an arbitrary radial scale, one can transform Eq. (\ref%
{radial}) into the following form (which is an exact transformation for $\mu
=0)$:%
\begin{equation}
\frac{d^{2}\mathrm{V}}{d\tau ^{2}}=-\frac{d\mathrm{V}}{d\tau }+2\sigma
\mathrm{V}^{3}.  \label{VV}
\end{equation}%
Formally, Eq. (\ref{VV}) is tantamount to the equation of motion of a
mechanical unit-mass particle with coordinate $\mathrm{V}$ and time $\tau $
in the normal ($\sigma =-1$) or inverted ($\sigma =1$) quartic potential, $W(%
\mathrm{V})=-\left( \sigma /2\right) \mathrm{V}^{4}$, under the action of
the friction force with coefficient $1$.

For $\sigma =-1$ (the focusing nonlinearity), an appropriate asymptotic
solution to Eq. (\ref{VV}) is one dominated by the balance of the potential
and friction forces at $\tau \rightarrow +\infty $, which is relevant for $%
r\ll r_{0}$, according to the definition of $\tau $ in Eq. (\ref{V}):
\begin{equation}
\mathrm{V}\approx (1/2)\tau ^{-1/2}.  \label{VVV}
\end{equation}%
For $\sigma =1$ (defocusing), the appropriate solution to Eq. (\ref{VV}) is
also determined by the balance of the potential force and friction, the
relevant region being $\tau \rightarrow -\infty $, i.e., $r\gg r_{0}$:%
\begin{equation}
\mathrm{V}\approx (1/2)\left( -\tau \right) ^{-1/2}.  \label{VVVV}
\end{equation}%
Both solutions (\ref{VVV}) and (\ref{VVVV}) eventually translate into the
asymptotic expression for the 3D density given above by Eq. (\ref{3D}).

In the opposite limit of $r\rightarrow \infty $, a straightforward
consideration yields an asymptotic expression for the solution in the form of%
\begin{equation}
\mathrm{U}_{\mathrm{3D}}(r)\approx Cr^{-1}\exp \left( -\sqrt{-2\mu }r\right)
,  \label{3Dexp}
\end{equation}%
with constant $C$, cf. Eq. (\ref{2Dexp}).

An exact scaling relation between the 3D norm and chemical potential, as it
follows from the cubic NLSE (\ref{u3D}), is the same as its counterpart (\ref%
{N2D}) in the 2D model with the quintic nonlinearity:%
\begin{equation}
N_{\mathrm{3D}}(\mu )\equiv 4\pi \int_{0}^{\infty }\mathrm{U}^{2}(r)r^{2}dr=%
\mathrm{const}\cdot \left( -\mu \right) ^{-1/2}.  \label{N3D}
\end{equation}%
This relation satisfies the anti-VK criterion, hence the family of the 3D
singular solitons may be (and indeed is) stable in the case of the
defocusing cubic nonlinearity, $\sigma =1$.

\subsubsection{Interpretation of the 3D singular solitons: screening of an
attractive spherical potential}

Similar to what is presented above for the 2D model, one can augment the 3D
model by the attractive delta-functional potential concentrated on a sphere
of radius $\rho $, the respective stationary equation being (cf. Eq. (\ref%
{2Deps}))%
\begin{equation}
\frac{1}{2}\left( \frac{d^{2}\mathrm{U}}{dr^{2}}+\frac{2}{r}\frac{d\mathrm{U}%
}{dr}\right) =-\mu \mathrm{U}+\mathrm{U}^{3}-\varepsilon _{\mathrm{3D}%
}\delta (r-\rho )\mathrm{U},  \label{3Deps}
\end{equation}%
where it is assumed that, although $\rho $ is small, it must be larger than $%
r_{0}$ in Eq. (\ref{V}). A stationary solution to Eq. (\ref{3Deps}) is
sought for in the form given by Eqs. (\ref{V}) and (\ref{VVVV}) at $r>\rho $%
, and as $\mathrm{U}=\mathrm{const}\approx \left( 2\rho \sqrt{\ln \left(
\rho /r_{0}\right) }\right) ^{-1}~~$at$~~r<\rho $. Then, calculations
similar to those outlined above in the 2D setting lead to a relation between
$\rho $ and the strength of the attractive potential, $\varepsilon _{\mathrm{%
3D}}=1/\left( 2\rho \right) $, cf. Eq. (\ref{eps2D}). In the limit of $\rho
\rightarrow 0$ (which is imposed along with $r_{0}\rightarrow 0$, so as to
keep condition $\rho >r_{0}$ valid), the respective \textquotedblleft
charge" of the 3D attractive potential is $Q_{\mathrm{3D}}=4\pi \rho
^{2}\cdot \varepsilon _{\mathrm{3D}}=2\pi \rho \rightarrow 0$, cf. Eq. (\ref%
{Q2D}) Thus, one may realize the 3D singular soliton as a state which
provides the screening of the vanishingly small ``charge".

\section{Numerical results for the 1D, 2D, and 3D singular solitons}

The numerical scheme for producing singular solitons as solutions of the
NLSEs with the self-repulsive nonlinearity must be adjusted to the fact
that, in the analytical form, the solutions take infinite values at the
origin. In Ref. \cite{HS}, a finite-difference scheme was used for this
purpose. It was defined on a grid with spacing $\Delta $, constructed so
that closest to the origin were points with coordinates
\begin{equation}
\left( x,y,z\right) =\left( \pm \Delta /2,\pm \Delta /2,\pm \Delta /2\right)
\label{Delta}
\end{equation}%
in the 3D case, and similarly in 1D and 2D. At these points, boundary
conditions with large but finite values of $|\psi |$ were fixed according to
the asymptotically exact analytical expressions (\ref{-1/3}), (\ref{2Drep}),
and ( \ref{3D}).

\begin{figure}[h]
\begin{center}
\includegraphics[height=5.5cm]{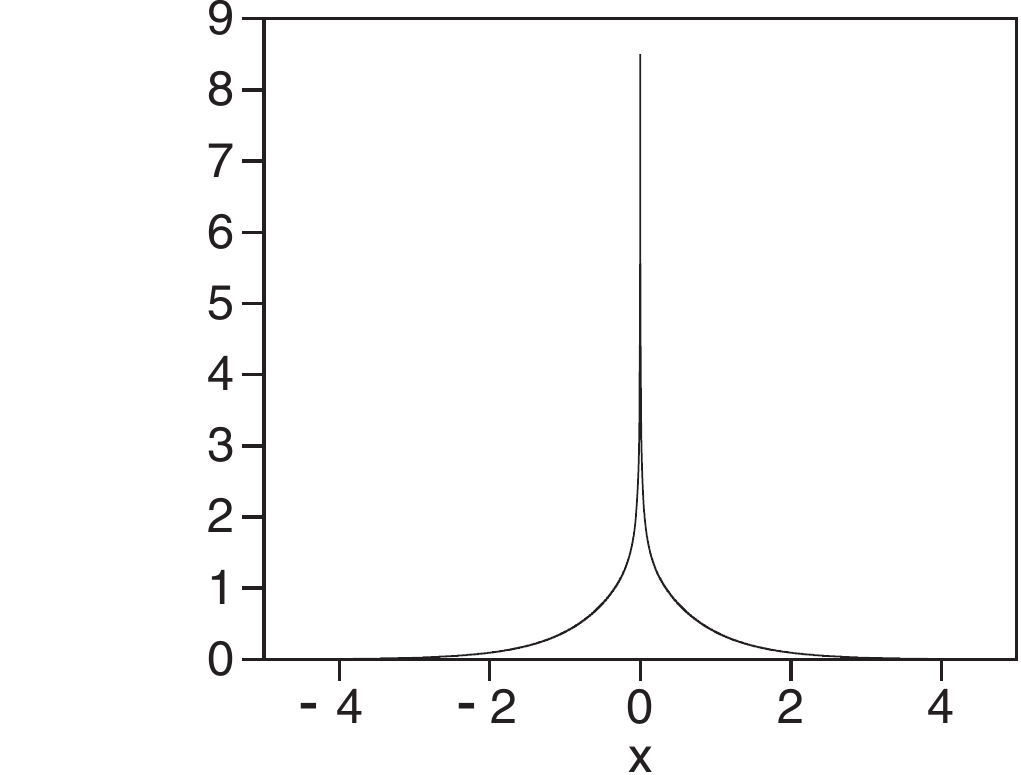}
\end{center}
\caption{The shape of the singular 1D soliton, as given by Eq. (\protect\ref%
{Uexact}) with $\protect\mu =-1$.}
\label{fig1b}
\end{figure}

A typical shape of the 1D singular soliton, produced by exact solution (\ref%
{Uexact}) with $\mu =-1$, is plotted in Fig. \ref{fig1b} with stepsize $%
\Delta x=10^{-5}$. In agreement with the fact that the entire family of the
1D singular solitons satisfies the anti-VK criterion, numerical results
corroborate the full stability of the family.

\begin{figure}[h]
\begin{center}
\includegraphics[height=7.0cm]{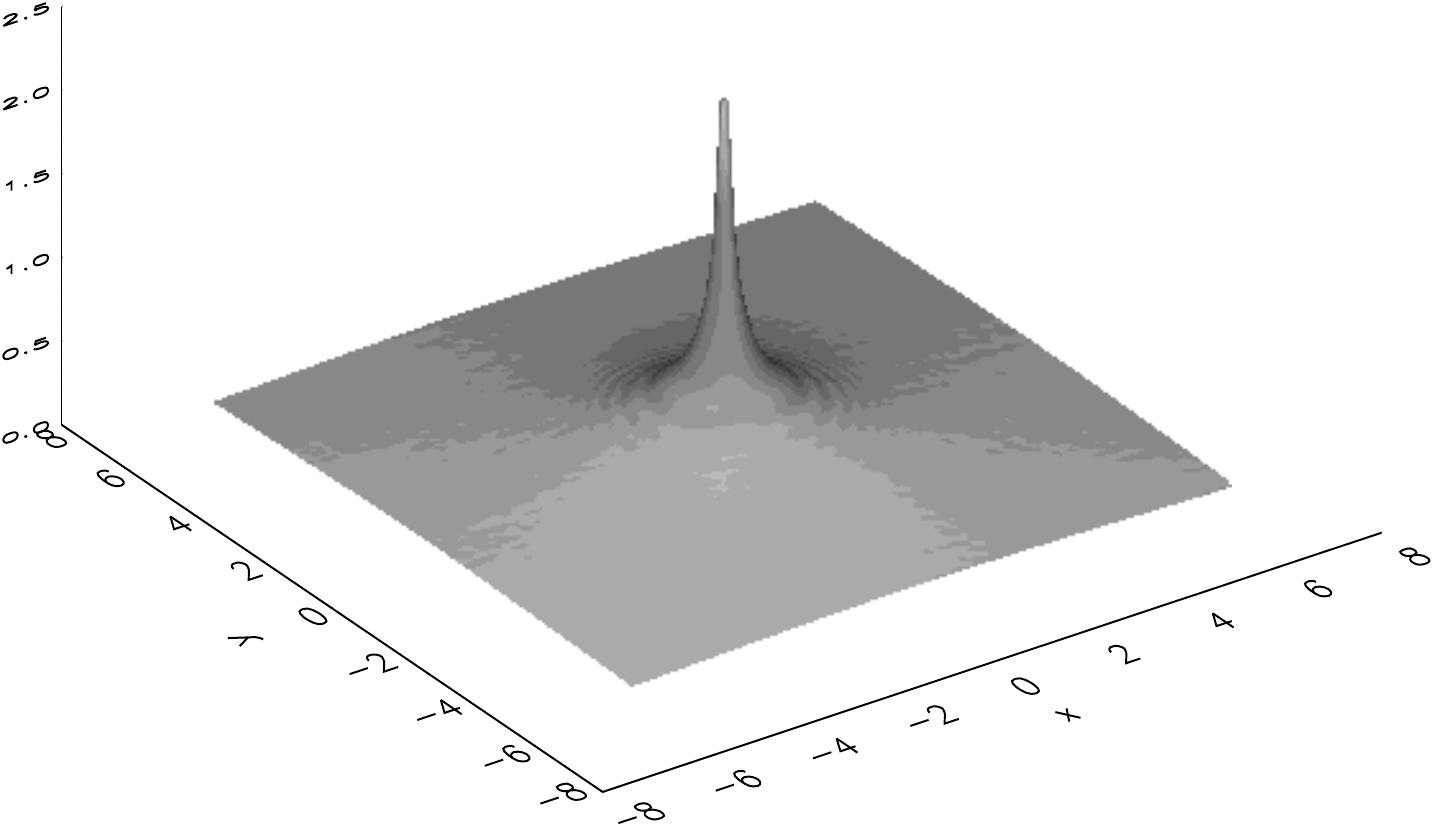}
\end{center}
\caption{A stable 2D singular soliton produced by simulation of the full 2D
equation~(\protect\ref{u2D}), starting from input (\protect\ref{u0}). }
\label{fig3b}
\end{figure}

In the 2D setting, numerical solution of the quintic equation (\ref{u2D})
has corroborated the existence and full stability of the singular solitons
with zero vorticity ($l=0$). As an illustration, Fig. \ref{fig3b} displays a
global view of the stable 2D soliton produced by direct simulations of Eq. (%
\ref{u2D}), starting from the initial conditions taken as per exact solution
(\ref{u0}) corresponding to $\mu =0$ (the formal divergence of its integral
norm at $r\rightarrow \infty $ is restricted by the finite size of the
integration domain). In particular, he numerical simulations confirm the
stability of the 2D singular solitons against azimuthal perturbations which
attempt to break the axial symmetry of the solitons.

\begin{figure}[h]
\begin{center}
\includegraphics[height=7.0cm]{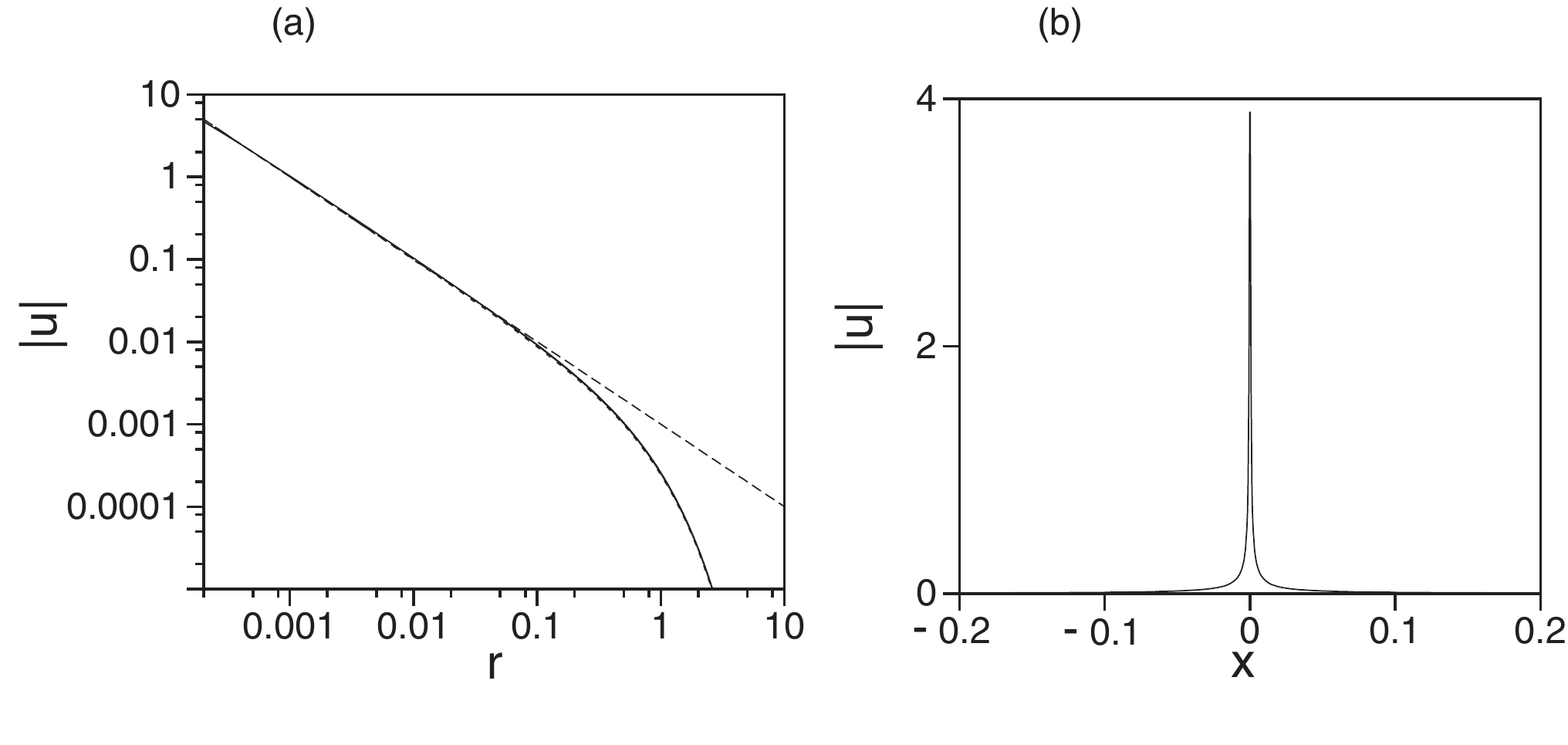}
\end{center}
\caption{(a) The comparison, on the double-logarithmic scale, of the
numerically found stationary solution of Eq.~(\protect\ref{u3D}) for the 3D
singular soliton with $\protect\mu =-1$ (the continuous line) with profile $%
0.001\cdot r^{-1}$ relevant at small $r$ (the long-dashed line), and the
global approximation provided by Eq. (\protect\ref{3Dexp}) with $C=0.001$
(the short-dashed line), see explanation in the text. The difference between
the latter analytical approximation and the numerical solution is barely
visible. (b) The radial cross section of the 3D singular soliton, produced
at $t=2$ by direct simulations of Eq.~(\protect\ref{u3D}), with the input
taken from panel (a). The result confirms stability of the 3D singular
soliton. In this figure, labels for $|\protect\psi |$ and $\mathrm{U}_{%
\mathrm{3D}}$ are replaced by $|u|$.}
\label{fig4b}
\end{figure}

In the 3D setting, numerical solution of Eq. (\ref{u3D}) produces a family
of 3D singular solitons. Comparison of these solutions with the analytical
prediction, given by Eqs. (\ref{V}) and (\ref{VVVV}) for small $r$, is not
straightforward, as the analytical expression contains indefinite parameter $%
r_{0}$. Therefore, an example, displayed in Fig. \ref{fig4b}(a) presents the
comparison of the numerical solution to analytical profile $\mathrm{const}%
\cdot r^{-1}$ at small $r$, with the constant selected as the best-fit
parameter ($\mathrm{const}=0.001$ for the case of $\mu =-1$, displayed in
Fig. \ref{fig4b}(a)).

Further, one can try to use the asymptotic form of the 3D solution at large $%
r$, given by Eq. (\ref{3Dexp}), with $C$ set equal to the best-fit value of $%
\mathrm{const}$ selected at $r\rightarrow 0$, as a global analytical
approximation. Figure \ref{fig4b}(a) demonstrates that such an approximation
is extremely close to its numerical counterpart at all values of $r$.

Finally, direct simulations of the evolution of the 3D singular solitons
confirm stability of the soliton family, see an example in Fig. \ref{fig4b}%
(b).

\section{Conclusion}

The aim of this paper is to produce a brief review of findings recently
reported in works \cite{HS} and \cite{we}, that demonstrate the existence of
several types of singular bound states in 3D, 2D, and 1D models with
self-repulsive nonlinearities. These states feature a singular structure of
the density at $r\rightarrow 0$, while the total norm converges, making the
states physically relevant solutions. One model combines the attractive
potential $\sim -1/r^{2}$ in the 2D space and the LHY (quartic)
self-repulsive term. The respective GPE (Gross-Pitaevskii equation) is a
model of BEC composed of particles carrying permanent electric dipole
moments, which are pulled to the central charge. Previously, it was found,
in the framework of the MF and many-body quantum settings alike \cite%
{HS1,Gregory}, that the quantum collapse, driven by this attractive
potential, can be effectively suppressed in the 3D space by the usual cubic
nonlinearity (which represents, as usual, repulsive interactions of
colliding particles), while the same cubic nonlinearity cannot stabilize the
2D condensate. A brief review of those results was presented in Ref. \cite%
{CM}.

The new result, reported in Ref. \cite{we} and summarized in Section 2 of
the present review, is that the quartic self-repulsive term is sufficient to
suppress the 2D quantum collapse and restore the missing GS (ground state),
which is a completely stable one.
States with embedded angular momentum (vorticity) are constructed too. They
are stable if the strength of the attractive potential exceeds a certain
threshold value, which depends on the vorticity.
An essential peculiarity is that stable 2D GS modes exist, counter-intuitively,
even when the attractive central potential is replaced by the repulsive one,
with a sufficiently small strength, as defined by Eq. (\ref{<4/9}). The
latter feature is closely related to another topic relevant to the
consideration of singular states, which was elaborated in Ref. \cite{HS} and
is summarized in Section 3 of the review. It deals with NLSEs (nonlinear
Schr\"{o}dinger equations) in 1D, 2D, and 3D free space, which contain the
septimal, quintic, and cubic repulsive terms, respectively. These equations
demonstrate the existence of stable singular solitons, that realize the
model's GS in all the cases, with the 1D solitons found in an exact
analytical form. The physical interpretation of these counter-intuitive
findings is provided by the possibility to construe the singular solitons as
a result of screening, by the respective self-repulsive nonlinearity, of a
delta-functional attractive potential, whose integral strength is divergent
in 1D, finite in 2D, and vanishingly small in 3D.

A common feature of both models considered in this paper is that, on the
contrary to the stable zero-vorticity GSs, two-dimensional vortex states are
completely unstable. Therefore, a challenging problem for further work is
search for physically relevant modifications of the 2D models which may
allow the existence of stable vortices. A still more challenging objective
is to construct 3D states with embedded vorticity and investigate their
stability. Another subject for further work may be singular dissipative
solitons in complex Ginzburg-Landau equations \cite{HS}. Quite challenging
may also be extension of the analysis for singular modes (if any) in models
of Fermi gases based on density-functional equations.




\noindent
\textbf{Acknowledgment}The work on topics relevant to the mini-review was supported, in part, by
the Israel Science Foundation through Grant No. 1286/17.





\end{document}